\documentclass[aps,prl,twocolumn,showpacs,superscriptaddress,groupedaddress,nofootinbib]{revtex4}  
\usepackage{graphicx}  
\usepackage{dcolumn}   
\usepackage{bm}        
\usepackage{amssymb}   
\usepackage{subfigure}
\usepackage{color}
\usepackage{tabularx}
\usepackage{multirow}
\usepackage{amsmath}
\usepackage{epstopdf}   

\usepackage{hyperref}

\begin{document}



\title{Low-mass dark matter search results \\ from full exposure of PandaX-I experiment}
\date{\today}
\vspace{0.5in}
\newcommand{\sjtuphys}{\affiliation{INPAC and Department of Physics and Astronomy, Shanghai Jiao Tong University, \\
Shanghai Laboratory for Particle Physics and Cosmology, Shanghai, 200240, P. R. China}}
\newcommand{\sjtume}{\affiliation{School of Mechanical Engineering, \\Shanghai Jiao Tong University, Shanghai, 200240, P. R. China}}
\newcommand{\sdu}{\affiliation{School of Physics and Key Laboratory of Particle Physics and Particle Irradiation (MOE), Shandong University, Jinan 250100, China}}
\newcommand{\sinap}{\affiliation{Shanghai Institute of Applied Physics, Chinese Academy of Sciences, Shanghai, 201800, P. R. China}}
\newcommand{\umich}{\affiliation{Department of Physics, University of Michigan, Ann Arbor, MI, 48109, USA}}
\newcommand{\pku}{\affiliation{School of Physics, Peking University, Beijing, 100080, P. R. China}}
\newcommand{\umd}{\affiliation{Department of Physics, University of Maryland, College Park, MD, 20742, USA}}
\newcommand{\yalong}{\affiliation{Yalong River Hydropower Development Company, Ltd., 288 Shuanglin Road, Chengdu, 610051, P. R. China}}

\sjtuphys
\sjtume
\sdu
\sinap
\umich
\pku
\umd
\yalong

\author{Xiang Xiao} \sjtuphys
\author{Xun Chen} \sjtuphys
\author{Andi Tan} \umd
\author{Yunhua Chen} \yalong
\author{Xiangyi Cui} \sjtuphys
\author{Deqing Fang} \sinap
\author{Changbo Fu} \sjtuphys
\author{Karl L. Giboni} \sjtuphys
\author{Haowei Gong} \sjtuphys
\author{Guodong Guo} \sjtuphys
\author{Ming He} \sjtuphys
\author{Xiangdong Ji} \thanks{Spokesperson: xdji@sjtu.edu.cn and xji@umd.edu} \sjtuphys\pku\umd
\author{Yonglin~Ju} \sjtume
\author{Siao Lei} \sjtuphys
\author{Shaoli Li} \sjtuphys
\author{Qing Lin} \thanks{Corresponding authors: \\xiaomengjiao@sjtu.edu.cn (Mengjiao Xiao); \\mcfate@sjtu.edu.cn (Qing Lin)} \sjtuphys
\author{Huaxuan Liu} \sjtume
\author{Jianglai Liu} \sjtuphys
\author{Xiang Liu} \sjtuphys
\author{Wolfgang Lorenzon} \umich
\author{Yugang Ma} \sinap
\author{Yajun Mao} \pku
\author{Kaixuan Ni}  \sjtuphys
\author{Kirill Pushkin} \sjtuphys\umich
\author{Xiangxiang Ren} \sdu
\author{Michael Schubnell} \umich
\author{Manbin Shen} \yalong
\author{Yuji Shi} \sjtuphys
\author{Scott Stephenson} \umich
\author{Hongwei Wang} \sinap
\author{Jiming Wang} \yalong
\author{Meng Wang} \sdu
\author{Siguang Wang} \pku
\author{Xuming Wang} \sjtuphys
\author{Zhou Wang} \sjtume
\author{Shiyong Wu} \yalong
\author{Mengjiao Xiao} \thanks{Corresponding authors: \\xiaomengjiao@sjtu.edu.cn (Mengjiao Xiao); \\mcfate@sjtu.edu.cn (Qing Lin)} \sjtuphys
\author{Pengwei Xie} \sjtuphys
\author{Binbin Yan} \sdu
\author{Yinghui You} \yalong
\author{Xionghui Zeng} \yalong
\author{Tao Zhang} \sjtuphys
\author{Li Zhao} \sjtuphys
\author{Xiaopeng Zhou} \pku
\author{Zhonghua Zhu} \yalong
\collaboration{The PandaX Collaboration}
\
\date{\today}

\begin{abstract}
We report the results of a weakly interacting massive particle (WIMP) dark matter
search using the full 80.1\;live-day
exposure of the first stage of the PandaX experiment (PandaX-I) located in the China Jin-Ping
Underground Laboratory. The PandaX-I detector has been optimized for detecting low-mass WIMPs, achieving
a photon detection efficiency of 9.6\%.
With a fiducial liquid xenon target mass of 54.0\,kg, no significant
excess events were found above the expected background.  A profile likelihood ratio analysis
confirms our earlier finding that the PandaX-I data disfavor all positive low-mass
WIMP signals reported in the literature under standard assumptions. A stringent bound on a low mass WIMP is set at WIMP mass below 10\,GeV/c$^2$, demonstrating that liquid xenon detectors
can be competitive for low-mass WIMP searches.

\end{abstract}

\maketitle

\section{Introduction}
The existence of gravitationally attractive ``dark matter''
that dominates the matter composition of the universe
has been firmly established based on overwhelming evidence
from astronomical and cosmological observations~\cite{dm:evidence}.
Whether such abundant matter consists of yet unknown elementary particles remains
one of the most pressing scientific questions.
There are strong theoretical motivations for the existence of beyond the Standard
Model physics, many of which naturally predict
new stable neutral particles at the electroweak symmetry breaking scale with weak interactions,
generically named weakly interacting massive particles (WIMPs)~\cite{ref:theoryoverview,Jung:1996}.
WIMPs are a leading dark matter (DM) candidate where weak interactions between
WIMPs and ordinary matter allow for a direct
search for these particles through particle physics experiments.
In recent decades, direct searches of WIMP interactions with terrestrial
detectors have been carried out in deep underground laboratories
worldwide with ever increasing discovery power~\cite{dm:expreview}.

Since 2008, a number of underground direct detection experiments have reported signals
that could be interpreted as WIMP interactions within the detector. Among those are the
DAMA/LIBRA experiment using NaI(Tl) crystals~\cite{ref:dama},
the CoGeNT experiment~\cite{ref:cogent} using point-contact
Ge detectors, the CRESST-II experiment~\cite{ref:cresst1} using
cryogenic CaWO$_4$ bolometers (excess not
reproduced in the recent experiment~\cite{ref:cresst2}),
as well as the CDMS-Si experiment using cryogenic Si bolometers~\cite{ref:cdms-si}.
Although the claimed signals are not generically consistent, they all point to
low to median WIMP mass in the range of 10
to 50\,GeV/c$^2$.
On the other hand, the ZEPLIN-III~\cite{ref:zeplin3}, XENON-100~\cite{xenon100_final}, LUX~\cite{ref:lux},
and PandaX-I~\cite{pandax:first}
experiments utilizing xenon, the DarkSide-50 experiment using argon~\cite{ref:darkside}, 
the SuperCDMS~\cite{ref:supercdms,ref:cdmslite} and CDEX~\cite{ref:cdex} experiments
using Ge as targets, as as well as the KIMS experiment~\cite{ref:kims} using CsI(Tl) 
crystals, are in disagreement with some or all of these claims.

To achieve sensitivities to WIMPs beyond the current experimental bounds, detectors with
larger targets, lower background, and lower energy threshold are required.
In the past decade, dual phase xenon detectors have rapidly emerged as one of the most
promising technologies in WIMP direct detection, leading
the WIMP search sensitivity in a wide range of
parameter space~\cite{xenon10,ref:zeplin3,xenon100_final,ref:lux},
demonstrating superior scalability
in mass, and the capability to shield against and reject background.
However, in comparison to the
cryogenic bolometers~\cite{ref:supercdms,ref:cresst2} or semi-conductor ionization
detectors~\cite{ref:cdmslite,ref:cogent,ref:cdex},
dual phase liquid xenon detectors have
not demonstrated the ability to obtain a comparably low energy threshold.
Conventionally, the issue is attributed to insufficient light collection efficiency
or a lack of understanding of the low energy nuclear recoil (NR) quenching factor.
In recent years, the LUX and PandaX collaborations operated newly designed liquid xenon detectors which were constructed to optimize light collection efficiency.
At the same time, a comprehensive model of
scintillation and ionization processes in xenon known as
the NEST~\cite{ref:nest, ref:nest0.98, ref:nest1.0},
developed with simple phenomenological models and based on consideration of
world data, is gradually being adopted in the xenon field.
The values of the relative scintillation efficiency (L$_{eff}$) from NEST decrease 
continuously down to zero energy, which is consistent and slightly lower than 
that from an independent phenomenological calculation~\cite{ref:mu}. 
These developments call for
careful re-examination of the low mass WIMP sensitivity using xenon detectors.
In Ref.~\cite{pandax:first}, we reported the first 17.4 live days null search results in
PandaX-I.
In this paper, we present an improved analysis including the full
PandaX-I data set, starting from May 26, 2014 to Oct. 16, 2014, with a total of
80.1 live-day exposure in the search for dark matter. We shall refer to these data as 
dark matter search data in the remainder of this paper.

\section{The PandaX-I experiment}
PandaX is a dual-phase liquid xenon dark matter experiment~\cite{pandax:tdr}
located at the China Jin-Ping
Underground Laboratory (CJPL)~\cite{ref:cjpl}.
The first phase PandaX-I is a pancake-shaped 120\,kg detector optimized
for light collection targeted for low mass WIMPs~\cite{pandax:tdr, pandax:first}.
The xenon chamber is a stainless steel inner vessel with an inner diameter
of 750\,mm, housing approximately 450\,kg of liquid xenon. The entire inner vessel
sits in an outer vacuum vessel constructed from 5-cm
thick high-purity oxygen-free copper serving also as a radon barrier and electromagnetic shield,
and enclosed by a passive shield made of copper, polyethylene, lead, and polyethylene,
from inner to outer layers. The gap between the outer vessel and the passive shield is
continuously flushed with boil-off nitrogen to maintain a radon level of less than
5\,Bq/m$^3$, more than a factor of 20 below the level in the experimental hall.
The central time-projection-chamber (TPC)
is a cylinder with diameter of 60\,cm and height of 15\,cm confined by
a cathode grid ($-15$\,kV) at the bottom, a gate grid ($-5$\,kV) and
an anode mesh (ground) separated by 8\,mm, below and above the liquid level
respectively, and a surrounding Polytetrafluoroethylene (PTFE)
reflective wall.
After a particle-xenon interaction, prompt scintillation photons (S1 signal)
are produced in the liquid. Ionized electrons are then drifted vertically
upward by an induced drift field and extracted into the gas by
an extraction field, producing the electroluminescence (S2 signal).
A top photomultiplier tube (PMT) array consists of 143 Hamamatsu R8520-406 (1-in square)
tubes and a bottom array holds 37 Hamamatsu R11410-MOD (3-in circular) tubes. The PMTs
view the active volume, collecting photons from both the S1 and S2 signals,
with the bottom array dominating the light collection for both S1 and S2 signals.
The radioactivity from the bottom PMT array is shielded by a layer of 5 cm thick 
LXe below the
cathode and the PMT surface.
The average dark rate per tube, i.e. the rate of random single photoelectrons (PEs), is
0.06\,kHz and 1.07\,kHz for top and bottom PMTs, respectively.
The time separation between
S1 to S2 signals gives the vertical position of the interaction, and the horizontal
position is encoded in the S2 charge pattern in the PMT arrays.
Multiple scatter events can be identified from the data by events which contain multiple S2 signals,
either separated in time if they happen at different vertical position
or separated in the horizontal plane if there are multiple charge clusters
in the PMT pattern. Gamma ray background produces electron recoil (ER) events
whereas the dark matter signal produces nuclear recoil events. The ratio of S1 and S2 signal area gives a powerful means of ER rejection when looking for DM-like NR signals~\cite{AprileDoke2010}.

The PMT waveforms,
amplified by a factor of 10 using Phillips 779 amplifiers,
are recorded by CAEN V1724
14-bit 100 MS/s digitizers.
The trigger for the data acquisition system (DAQ) is generated based on the “majority”
outputs from the five digitizer boards for the bottom PMT arrays.
For low energy signals in the dark matter region,
the trigger is generated by S2 with a threshold of about 89 total
PE, whereas higher energy events were triggered
primarily by S1 with a charge threshold of about 65 PE.
Each readout
window is 200\,$\mu$s long, with approximately equal division of pre- and
post-trigger readout times. The PMTs are balanced to a gain of $2\times10^6$,
with a recorded
amplitude of the single photoelectrons roughly at 60 digitizer bits. To save data volume,
segments with waveform samples less than 20 digitizer bits from a pre-loaded baseline
are zero-suppressed. For non-suppressed segments, 40 time samples before and
after the 20 bit threshold crossing are recorded.
%

Three types of data runs were taken during the PandaX-I running period, the
WIMP search, $^{60}$Co ER calibration, and $^{252}$Cf
NR calibration runs.
A summary of the data taken is given
in Table~\ref{tab:data}. Various cuts (discussed below)
are applied to remove periods with unstable operating conditions,
leading to a difference between DAQ time and the live time.
\begin{table}[!htbp]
\centering
  \begin{tabularx}{0.45\textwidth}{crrr}
    \hline
    \multirow{2}{*}{Run type }&~~ DAQ Time &~~ Live Time &~~ Trigger Rate \\
    & (hr) & (hr) & (Hz) \\
    \hline
    DM & 2,158.32 & 1,923.11 & 3.58        \\
    $^{252}$Cf  & 95.32  & 94.05  & 17.95        \\
    $^{60}$Co   & 405.14 & 361.47 & 22.23       \\
    \hline
  \end{tabularx}
  \caption{Summary of data taken during the entire PandaX-I running period.}
  \label{tab:data}
\end{table}

Two independent analyses were developed within the collaboration,
utilizing different signal window selection, signal identification and
reconstruction, event selection cuts and efficiencies, as well as the final fitting method.
The two analyses were thoroughly cross checked at various analysis stages, yielding
consistent results. In the remainder of this paper, we will elaborate one of the
analyses, and the other one is detailed in Ref.~\cite{ref:lq_thesis}.

\section{Data processing and selection cuts}
A number of improvements have been made in the data analysis pipeline compared with the
first results~\cite{pandax:first}. We shall
describe the general procedure in steps below, with major improvements highlighted.

The raw data files are screened for basic data quality before being processed
for physics analysis. Detection of PMT high voltage outages is applied to filter data sets with low light collection. A nominal trigger rate below 10\,Hz is required to
reject those data sets which are seriously contaminated by noise during times when running conditions are poor.
Files with unexpected discharges from electrodes can be discriminated using the average
number of S1-like and S2-like signals in a waveform. If containing an average of larger
than 40 S1-like or 10 S2-like signals, the events will be removed. Dark rates from PMTs are tracked and used to
characterize the stability of the detector. A low random coincidence rate is
essential and a cut is developed on PMT dark rates to minimize contamination.

Baseline subtraction is performed on each waveform.
In this analysis, the baseline is calculated based on the pre-samples
from each waveform segment to suppress the drift and overshoot
of baselines, whereas in Ref.~\cite{pandax:first} only
weekly calibrated baselines were loaded. This update caused a
downward shift of the light yield of approximately 6\% at
40\,keV$_{ee}$ electron-equivalent energy.

Several malfunctioning PMTs are inhibited in the analysis.
During operation, four bottom PMT channels gradually developed
connection problems, manifest as improper base resistance or capacitance,
and were inhibited in the analysis to avoid a time dependent light yield.
Among the rest of the bottom PMTs, a number of them experienced
excessive dark rate (10\,kHz and above) during the run
but could sometimes be recovered
through power cycling or lowering the corresponding high voltage.
One channel was fully inhibited due to unstable dark rate.
The channel inhibition led to another 10\%
reduction in light yield. On average, two to three
bottom PMTs had to run at a lower gain ($<1\times10^{6}$) to maintain a manageable
dark rate. For the top array, seven PMTs gradually developed problems during the run and were
inhibited as the problem showed up, but has lesser impact to the analysis presented here.

Gain correction was applied to baseline-subtracted waveforms based on the
the results of weekly LED calibration runs. A hit finder algorithm identifies signal hits channel-by-channel while tagging noise primarily due to the periodic 200\,kHz electromagnetic 
interference
from the CAEN PMT high voltage supplies, occasionally fluctuated above the 
20 bit zero suppression threshold.
The waveform is then integrated in the ``hit window'' to define the hit charge.
The hits in different channels are then clustered in time using an improved
charge dependent algorithm with high efficiency for in-time short S1
signals while avoiding splitting a low charge but
wide S2 into multiple clusters.
For each cluster, a software sum is formed on all digitized channels,
from which one computes the
full-width-half-maximum, the full-width-1/10-maximum, as well
as the number of “peaks” in the cluster.
A binary decision tree method
is developed to sort any given cluster into S1-like and S2-like signals based on
these variables. The identification efficiency is verified to be nearly 100\%
by checking waveforms of thousands identified clusters by eye.

We developed further signal-level cuts to identify spurious noise in the
S1-like and S2-like signals.
For S1-like signals, we employed a further ripple-pattern cut on the software summed
waveform, a cut on the ratio of charge computed from the summed waveform 
to the total hit charge~\footnote{The summed waveforms for the 200\,kHz noise
tend to show a clear ripple feature, leading to a cancellation in the corresponding
charge.}, and a cut on the ratio of the total number of noise hits 
to the total hits in the
cluster. In addition, cuts are placed on the ratio of the height to area and
that of the height to width. To avoid signals from afterpulsing, S1 signals are required
to be before the first good S2 signal.
For S2, we developed a shape symmetry cut to
remove events very close to the anode when S1 and S2 cannot be easily separated
in time, identifying those events with a characteristic sharp spike at the beginning of the
S2 pulse.
In addition, S2-like signals will be discarded if its ratio to the
largest S2 is less than 1\% or if such signal is consistent with a single-electron
S2 with charge less than 30 PE; the inefficiency due to these two cuts is estimated
to be negligible by a Monte Carlo (MC) simulation.

Periods with unstable overall
PMT signal rates were identified during the run, possibly due to
small discharges on the electrodes, producing light pulses. In addition to the
earlier file-by-file
cuts, a number of tighter data quality
cuts were placed event-by-event to search for ``dirty waveforms'', including a cut
on the total number of S1-like signals, the ratio of S1 to total charge before
the first S2, and the ratio of the sum of S1 and S2 to the total 
charge (which also suppresses multiple scattering events). To
avoid ambiguity due to multiple S1-like signals per event, we require that
the number of S1
should be either 1 or 2, and in case of the latter the maximum
S1 is identified as the primary if the charge of the second S1 does not exceed
50\% of the first. Finally, to suppress accidental background, we placed a
3-hit coincidence cut on any good signal, and a 300 PE cut on S2 (discussed later).

Finally, S1 and S2 signals are reconstructed into physical events.
The vertical position is determined by the separation between S1
and S2 signals, assuming a drift speed of 1.7\,mm/$\mu$s under the drift field of
0.67\,kV/cm~\cite{Yoshino:76,pandax:first}.
Horizontal position of the interaction is reconstructed with the S2 PMT charge pattern using multiple algorithms. As an improvement to the
charge center of gravity (CoG) method,
we developed a fast charge pattern template matching (TM) method. The
expected charge templates were generated
using a custom GEANT4~\cite{ref:G4} based Monte Carlo which simulates
optical photon propagation from S2 signals in the PandaX-I TPC geometry,
identical to the templates used in the fast artificial neural network (FANN) reconstruction
method developed by the independent analysis.
The difference in the horizontal positions from the FANN and TM methods is
on average 5\,mm, obtained using 40\,keV de-excitation events from neutron
calibration data. This is independent of the radial and vertical positions and consistent with
the expected position resolution from MC.
In the analysis presented here, the TM reconstruction is chosen. To identify
multiple scattering at the same vertical location, we set a
charge clustering cut by requiring the horizontal distance between the CoG and TM
positions be less than 55\,mm apart.


\section{Detector Calibrations}
\label{sec:calib}

The PandaX-I detector has been carefully calibrated using various methods
to perform an effective search for low-mass WIMPs. Single electron
events were identified to calibrate the single electron gain (SEG) in
the electroluminescence. Neutron-activated X-rays were used to determine 
the photon detection efficiency for S1 (PDE) of the PMTs which signifies the sensitivity
of our detector in the low-mass region, and the electron extraction efficiency (EEE) from the liquid.
A neutron calibration with $^{252}$Cf was used to generate the NR events that were
used to define the DM search window.
Finally, a gamma calibration with $^{60}$Co was used to find the leakage of
ER background into the search window.

Neutron calibrations have been taken several times throughout the run with
a total exposure of 95 hrs (see Table I).
The 40\,keV ($^{129}$Xe) and 80\,keV ($^{131}$Xe) inelastic recoil X-rays are used in
calibrating the uniformity for both S1 and S2 in the detector. For a fixed energy
deposition in the detector, the PMT arrays see different light and charge yields depending on the
spatial location of the event, which must be corrected to a detector average before further analysis
is performed.
The uniformities for S1 and S2 are verified to be decoupled in the vertical direction
and horizontal plane.
The horizontal variation of 40\,keV S2 peaks in the 54.0 kg fiducial volume
is measured to be $\pm$36\%, which dominates the detector
non-uniformity. The vertical uniformity for the S2 signals, characterized
by an exponential ``electron lifetime'', reflects the electronegative
impurity level in the detector which tends to attenuate the charge signal during drifting.
The average electron lifetime in our detector is fit with a decaying exponential and
determined to be
328$\pm$8\,$\mu$s (an attenuation length of about 60\,cm as
compared to the 15\,cm maximum drift distance).
On the other hand, the variation of the S1 peak in the fiducial volume
in the vertical (horizontal) direction is $\pm$8.5\% ($\pm$9.5\%).
All discussions in the remainder of the paper
are made with the uniformity corrections taken into account.

\begin{figure}[!htbp]
  \centering
  \includegraphics[width=0.49\textwidth]{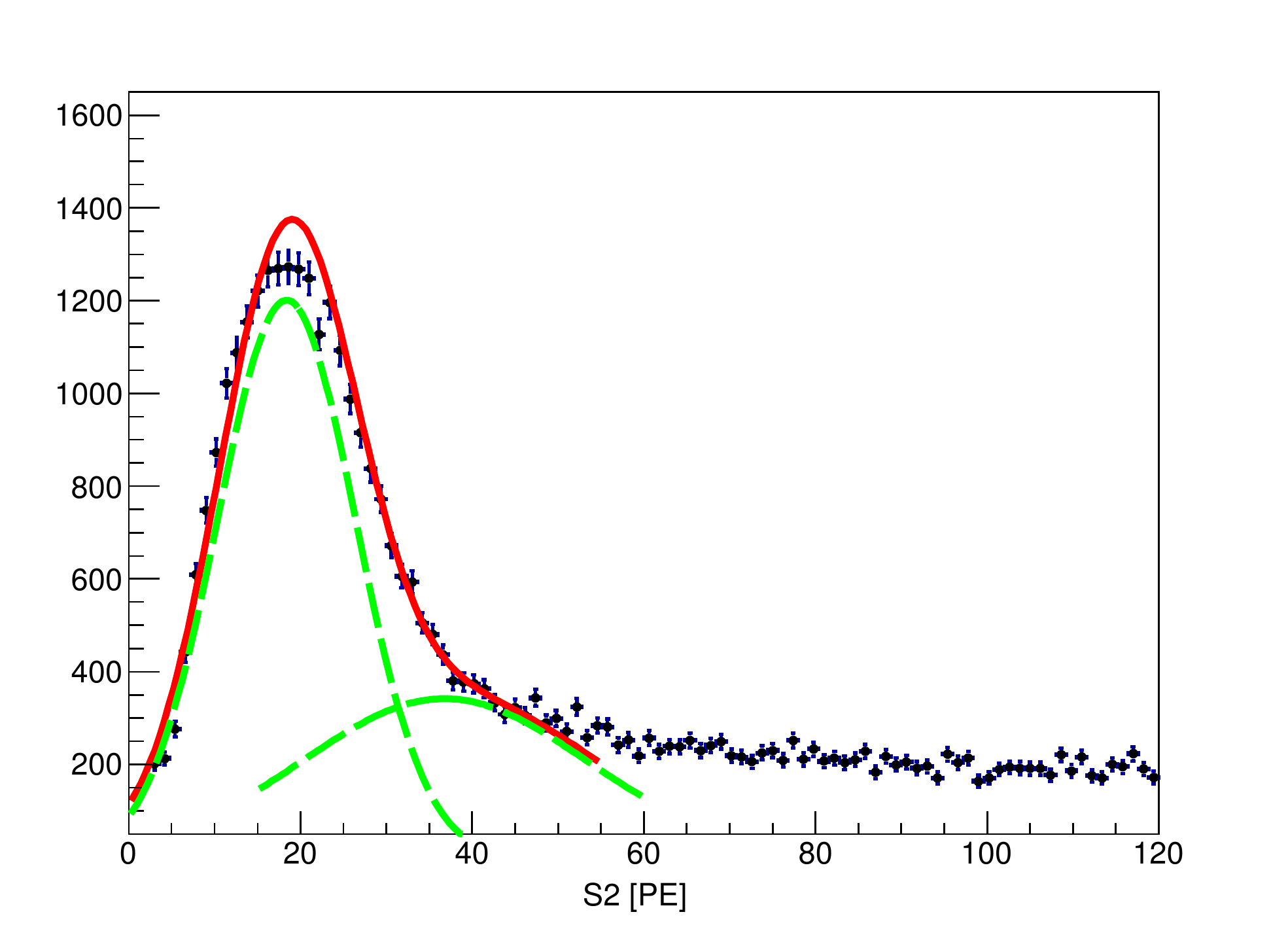}
  \caption{The PE distribution of the smallest S2 signals (corrected for
    horizontal non-uniformity), summed over all
    top and bottom PMTs. The single electron gain is determined by fitting 
    two constrained Gaussians, shown as the dashed green lines.}
  \label{fig:gasGain}
\end{figure}

One of the important properties of the detector is SEG,
the average number of PEs observed in PMTs from single-electron electroluminescence. It
can be determined from the PE distribution of the smallest S2 signals taken at any
normal detector run, fit
with a double Gaussian function with means related by a factor of 2 from the charge quantization (see Fig.~\ref{fig:gasGain}).
The SEG is determined to be 18.4$\pm$1.6 PE/e, where the uncertainty
is estimated by varying parameters in the hit clustering algorithm as well as the fitting function and range.

\begin{figure}[!htbp]
  \centering
  \includegraphics[width=0.49\textwidth]{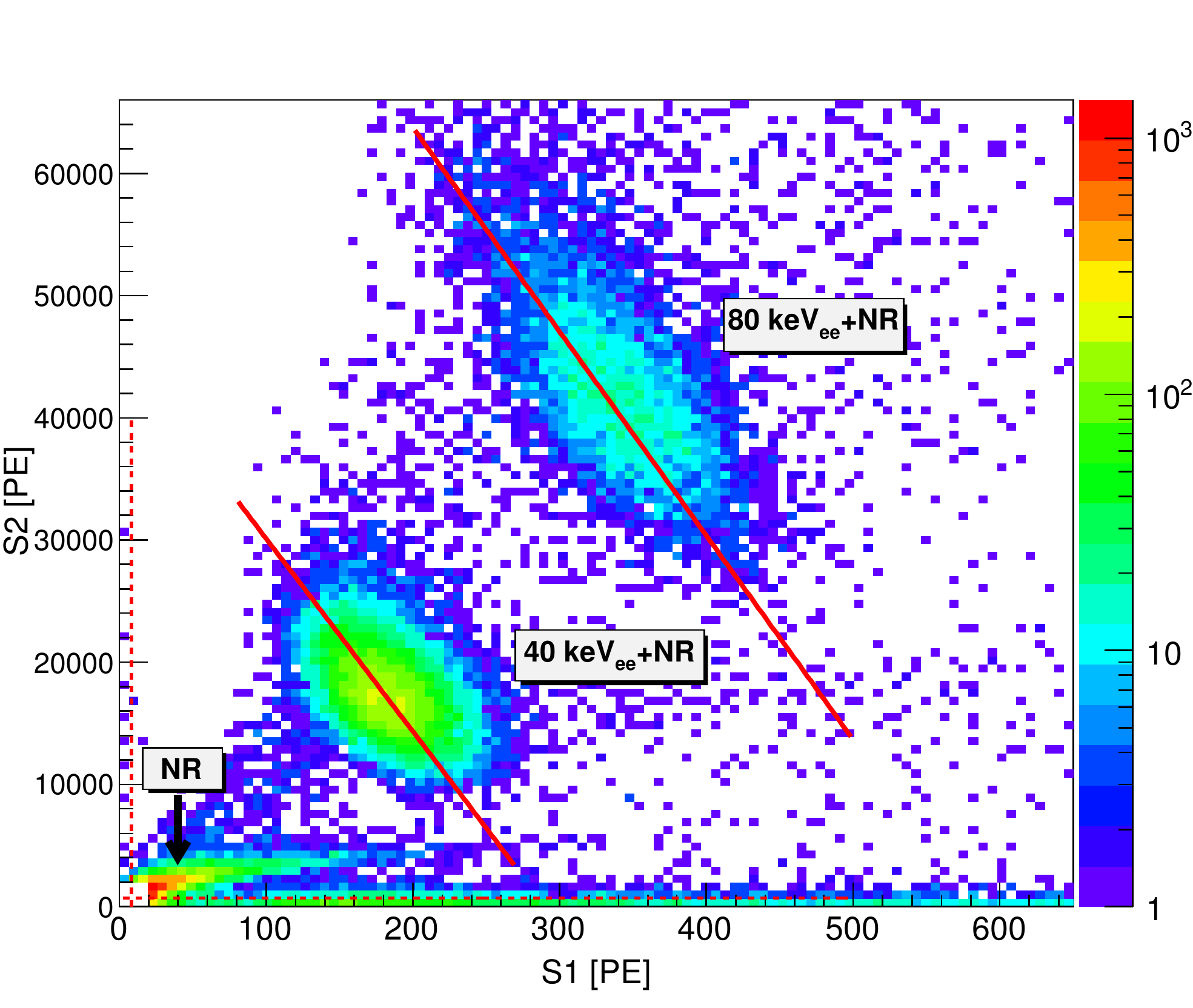}
  \caption{Distribution in uniformity-corrected S1 and S2 for the de-excitation gamma peaks
    in the neutron calibration runs. The anti-diagonal lines are the anti-correlation fit
    at the two energies. The dashed vertical and horizontal lines indicate the mean
    NR energy in mixture with the gamma energies at 40 and 80 keV.}
  \label{fig:anti}
\end{figure}

The PDE and EEE can be determined from
the 40 and 80\,keV X-ray events during the neutron calibration.
The events collected are shown in the S2 vs. S1 plot in Fig.~\ref{fig:anti}.
The location of the 40\,keV
peak (with decay time  less than 1\,ns) is at 178.8 PE in S1, with an average 11.6 PE mixture
from the associated NR, estimated from the pure NR events seen at low energy
as well as through MC. At this energy, our detector has a S1 photon yield of 4.2 PE/keV. 
Using the NEST-0.98 model~\cite{ref:nest0.98}, this corresponds to 6.0 PE/keV 
at zero electric field at the standard 122\,keV, in comparison to the 3.9 PE/keV obtained 
XENON100~\cite{xenon100_final} and
8.8 PE/keV in LUX~\cite{ref:lux}.

The electron-equivalent energy of a given events can be 
reconstructed from the light and charge outputs as
\begin{equation}
E_{rec} = {\rm (S1/PDE  + S2/SEG/EEE)} \times W,
\label{eq:comb_energy}
\end{equation}
%
where $E_{rec}$ is the reconstructed energy in keV$_{ee}$ splitting
into scintillation and ionization parts, and  $W =13.7$\,eV is the average energy
to produce a scintillation photon or to liberate an electron~\cite{ref:nest}.
The anti-correlated fluctuations in the light and charge outputs due to electron-ion
recombination is naturally accounted for in Eq.~\ref{eq:comb_energy}.
Similar to Ref.~\cite{pandax:first}, we performed
anti-correlation fits using Eq.~\ref{eq:comb_energy} to the 40 and 80\,keV de-excitation
peaks, as well as the neutron-induced meta-stable
$^{129m}$Xe (164\,keV)  decay gamma rays
after the neutron calibrations~\footnote{We did not perform anti-correlation fits for the
$^{131m}$Xe 236\,keV gamma lines since it was difficult to separate the peak cleanly
from the background.}.
The PDE (EEE) determined with the 40\,keV$_{ee}$
peak is 9.6\% (82.1\%). The fractional uncertainties are estimated
to be 10\% and 9\%, respectively, based on the difference in values obtained at
other two energies, as well as those in Ref.~\cite{pandax:first}.



\begin{figure}[!htbp]
  \centering
  \includegraphics[width=0.49\textwidth]{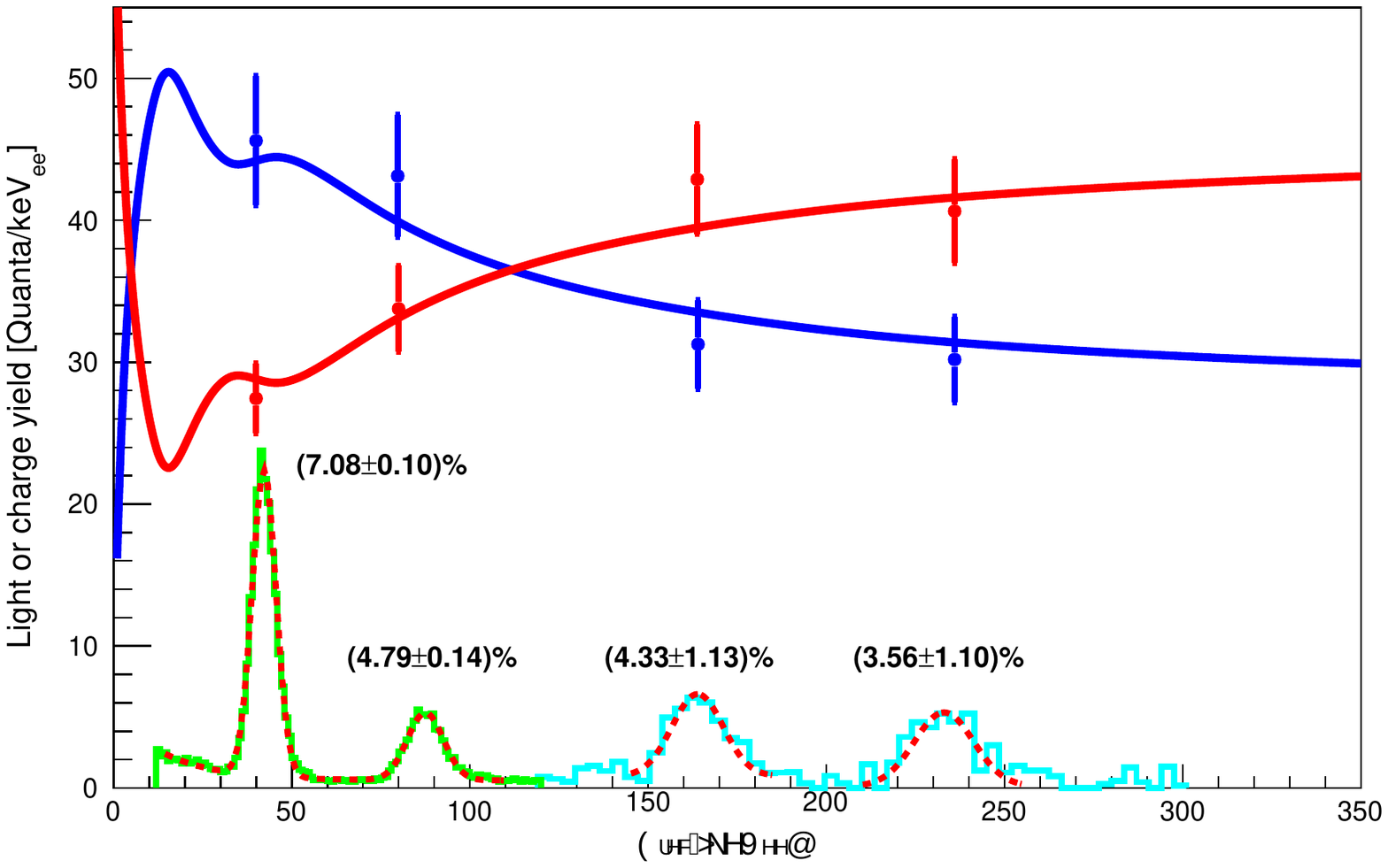}
  \caption{
    The $L_y$ (blue) and $C_y$ (red) (in units of quanta per keV$_{ee}$)
    extracted
    based on PDE and EEE obtained at different energies overlaid with
    corresponding curves predicted by NEST-0.98.
    The reconstructed
    energy spectra for the de-excitation peaks and meta-stable xenon isotopes
    are also overlaid with y axis scaled for visual clarity with fitted energy resolutions
    indicated in the figure.
  }
  \label{fig:eng_model}
\end{figure}

To facilitate the comparison of our data with model prediction, we
convert the peaks in S1 and S2 into a per unit energy total photon yield ($L_y$)
and charge yield ($C_y$), using
\begin{eqnarray}
\label{eq:Ly_Cy}
L_y =& \langle S1 \rangle /{\rm PDE}/E_{rec}, \nonumber\\
C_y =& \langle S2 \rangle /{\rm SEG/EEE}/E_{rec}\,,
\end{eqnarray}
where $\langle S1 \rangle$ and $\langle S2 \rangle$ here refer to the location of corresponding
peaks in the distribution. In Fig.~\ref{fig:eng_model}, our measured data is compared
to the mean values
in NEST-0.98~\cite{ref:nest0.98} under the same drift field. Reasonable agreement
is found at all four energy peaks in $^{252}$Cf data (40, 80, 164, 236 keV). The uncertainties
shown in the figure, aside from the statistical uncertainties in the peak determinations, arise from
the systematic uncertainties of the PDE and EEE determination
through the anti-correlation fits.

\begin{figure}[!htbp]
\centering
 \includegraphics[width=0.49\textwidth]{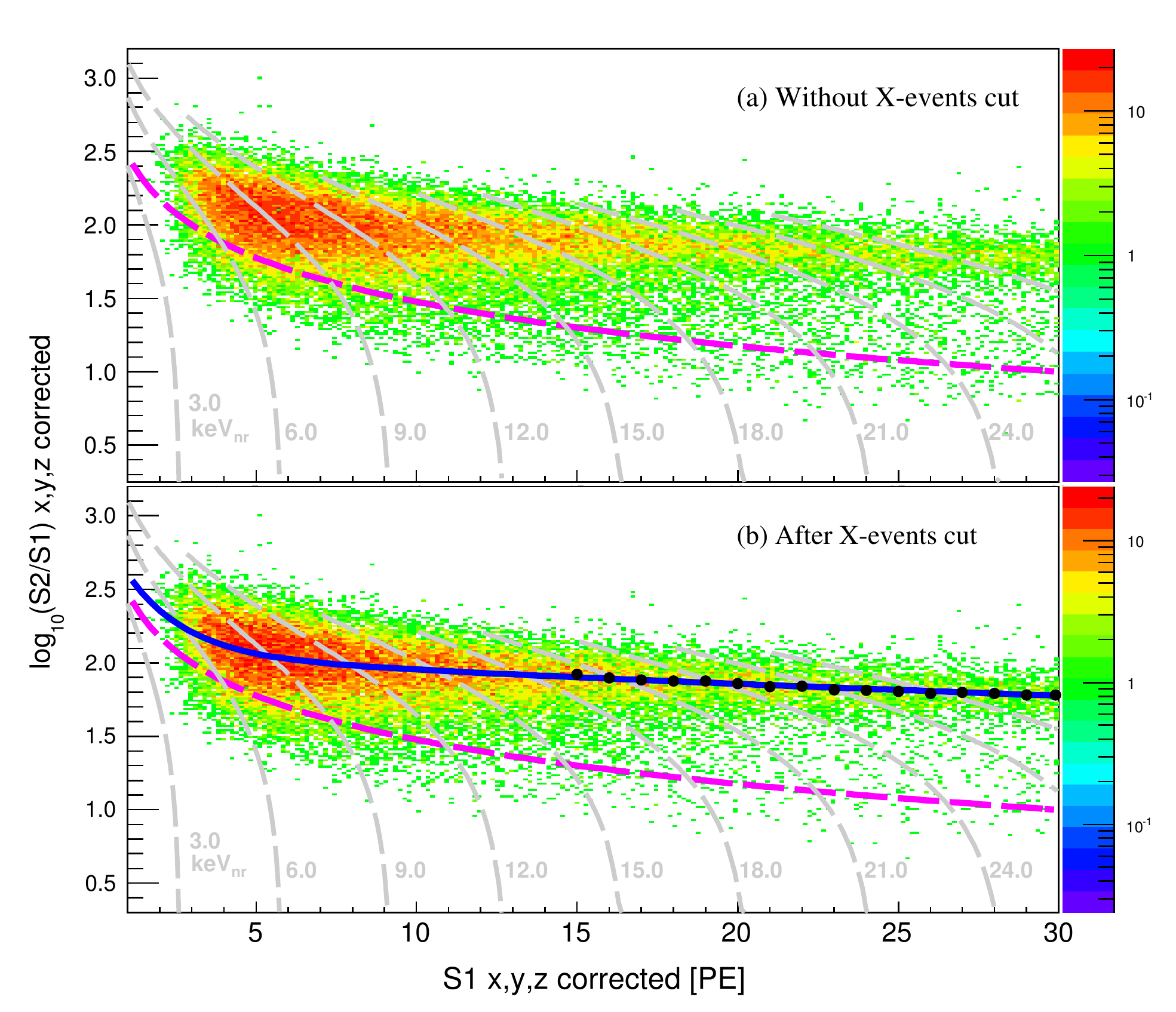}
\caption{The band of log$_{10}({\rm S2}/{\rm S1})$ versus S1 for the NR calibration data
without (a) and with (b) the ``X'' cut. See text for definition of the cut.
The 54.0-kg fiducial cut is applied. The solid blue line in (b) is the
median of the pure NR band in MC, and the dots are the Gaussian mean obtained from the
data for S1$>15$ (where the detection efficiency is flat so that 
the data and MC comparison
is straightforward). The dashed magenta lines in both figures are the
300 PE cut on S2, below which no dark matter candidate is considered. The gray
dashed lines are the equal energy lines with NR energy indicated 
in the figures.
}
\label{fig:NR_band}
\end{figure}

In the $^{252}$Cf NR calibration runs, the single events at very
low energy with S1$<30$ PE are expected to have less than 1\% contamination from the ER
band based on MC simulations, and the latter can therefore be neglected.
The distribution of these low energy events in log$_{10}$(S2/S1) vs. S1 is shown in
Fig.~\ref{fig:NR_band}(a).
It can be seen that there are scattered events with suppressed S2, producing an
asymmetric NR band.
Based on the charge pattern of S1 signals,
it was determined that such events (called ``X'' events~\cite{xenon10_0}) are due to
multiple scattering of neutrons with some energy deposition
in the ``chargeless'' region, either below the cathode, or in the xenon ``skin''
between the PTFE wall and the stainless steel inner
vessel (viewed partially by the outermost ring of the top PMT array). They have to be
properly taken into account to correctly calibrate the NR efficiency.

\begin{figure}[!htbp]
\centering
\includegraphics[width=0.5\textwidth]{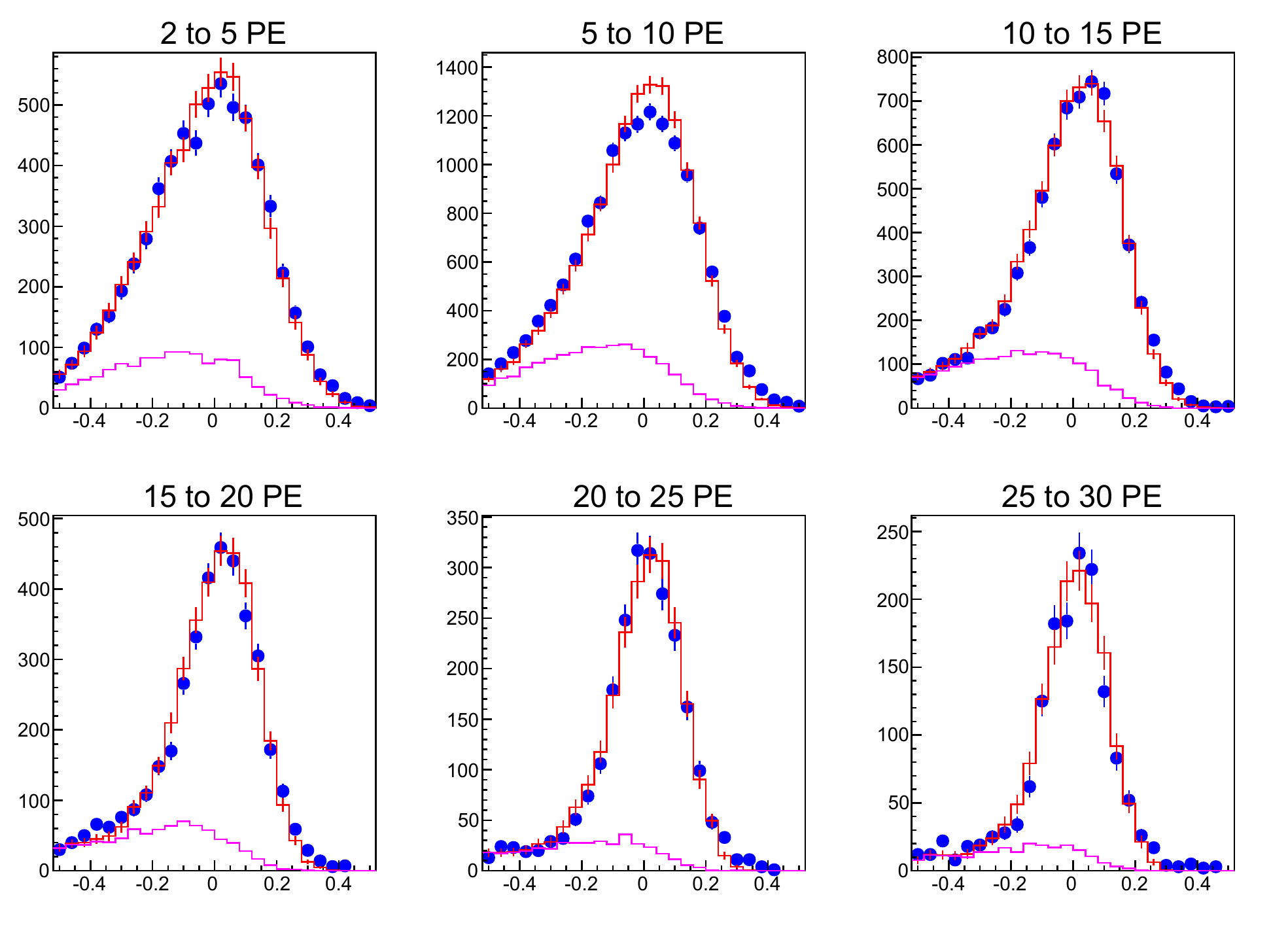}
\caption{Comparison of the distribution of $\log_{10}({\rm S2}/{\rm S1})$ in the
NR data and tuned MC
in six slices of S1 as indicated by the figure titles: data (blue),
tuned ``neutron-X'' events in the MC (magenta), and the sum of pure NR and 
tuned ``neutron-X'' in MC (red).
In each slice of S1, the value of $\log_{10}({\rm S2}/{\rm S1})$
is shifted relative to the median value in that slice.
The efficiency in Fig.~\ref{fig:NR_eff} has been
applied to the MC to compare with the data.
}
\label{fig:s1_slice_comp}
\end{figure}

To compare the data with expectation, a GEANT4 MC is developed
to simulate the $^{252}$Cf runs, which produces both single-scatter
pure NR and ``neutron-X''
recoil spectra, and employs the NEST-0.98 nuclear recoil model~\cite{ref:nest0.98} with the
PDE, EEE and the SEG obtained above.
After global tuning of the strength of the "neutron-X" events in the MC, 
excellent agreement is found in different slices of S1 
between the data and MC (Fig.~\ref{fig:s1_slice_comp})~\footnote{A fluctuation of 17\% from the gas gain, in addition to
the nominal statistical fluctuations introduced by NEST, helps to match
the measured width in each S1 slice.}.
If the new NEST-1.0 model~\cite{ref:nest1.0} is used instead, the MC can also be tuned to agree with the
data by increasing the EEE up by 2\%, much less than its assigned uncertainty.
The tuned MC is used as the true physical distribution to
extract the NR efficiency.

\begin{figure}[!htbp]
  \centering
  \includegraphics[width=0.49\textwidth]{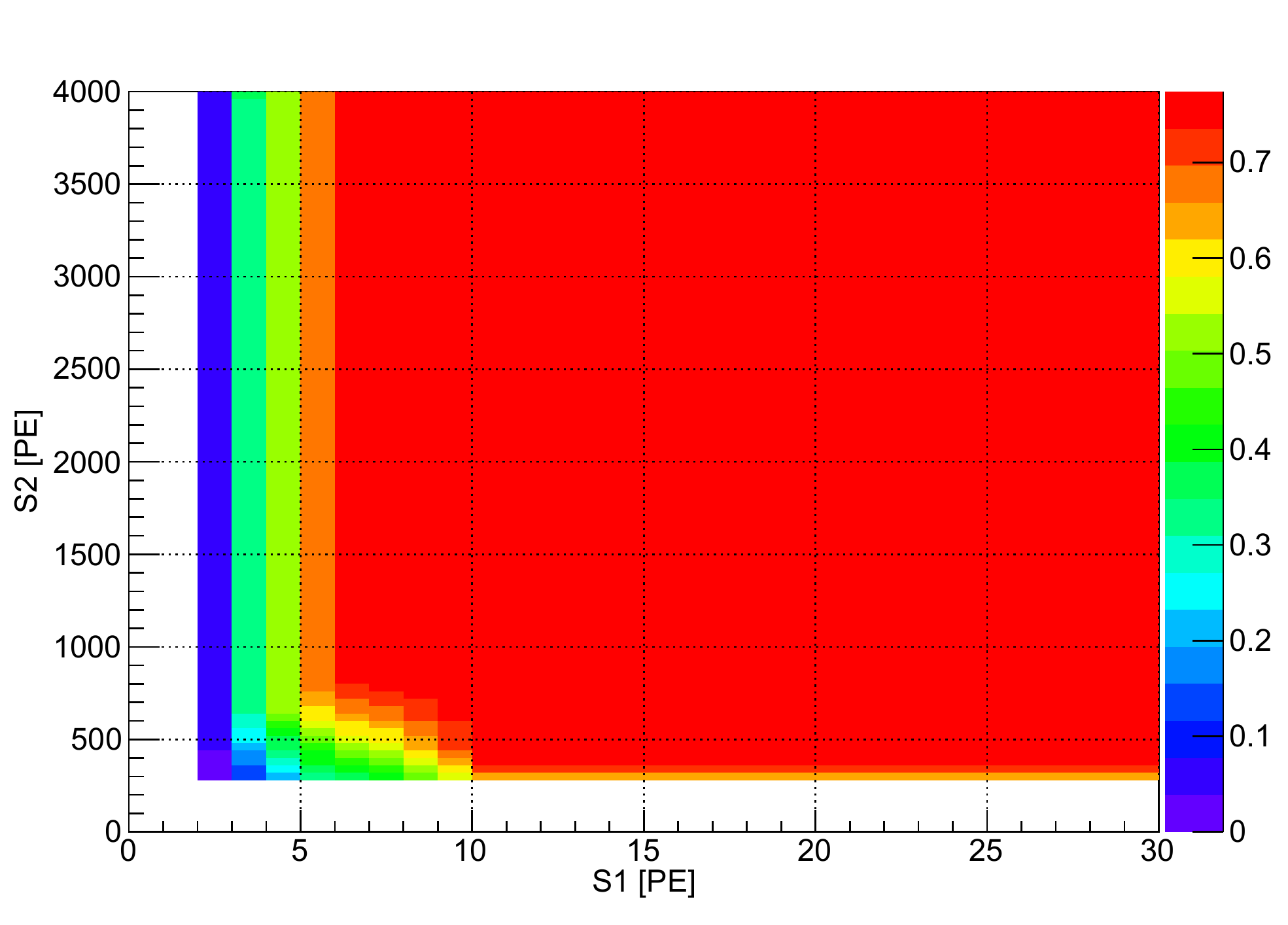}
  \caption{
    The nuclear recoil efficiency in (S1,S2) determined using the method
    discussed in the text.
  }
  \label{fig:NR_eff}
\end{figure}

To suppress the ``X'' events, a charge asymmetry cut between the top and bottom
PMT arrays as well as a cut on the ratio of the maximum
single PMT charge to the total on S1 were applied to all data including
$^{252}$Cf, $^{60}$Co, and DM data sets.  The NR distribution after the cut
is shown in Fig.~\ref{fig:NR_band}b, where
the low S2 ``X'' events are significantly reduced.
Our overall analysis cut
efficiency for NR events with S1\,$>10$~PE is estimated by comparing the number of
$^{252}$Cf events in (S1,S2) bins
before and after all cuts in this energy region,
and an approximately uniform 77.5\% value
is obtained. At lower energy, the overall NR efficiency is estimated by
taking the ratio of the measured distribution to the tuned MC,
anchored at 77.5\% at higher energy. The
resulting two-dimensional distribution of the NR efficiency is shown
in Fig.~\ref{fig:NR_eff}.

\begin{figure}[!htbp]
  \centering
  \subfigure[ER band]
  {
    \label{fig:ER}
    \includegraphics[width=0.49\textwidth]{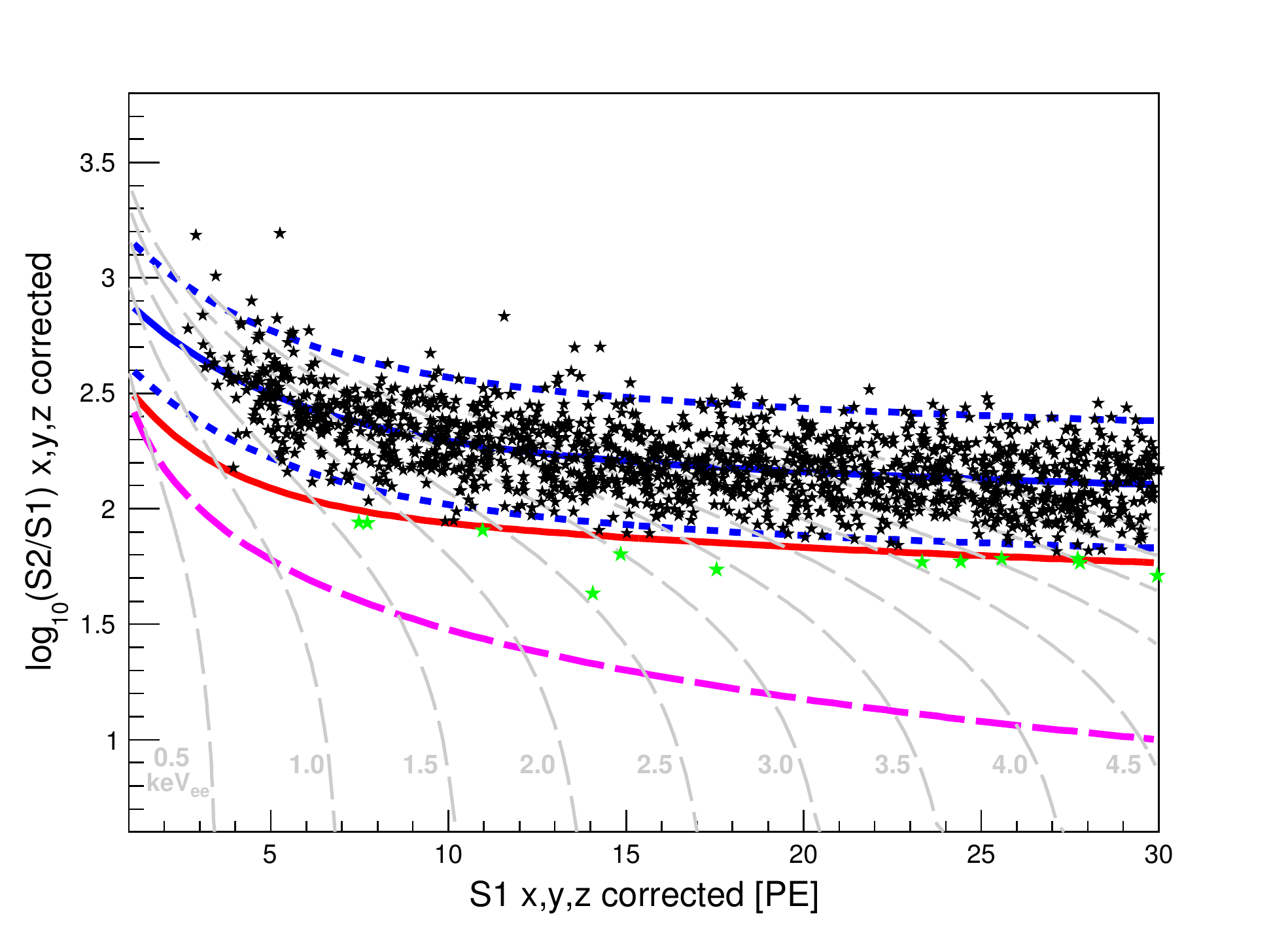}
  }
  \subfigure[ER efficiency]
  {
    \label{fig:ER_eff}
    \includegraphics[width=0.49\textwidth]{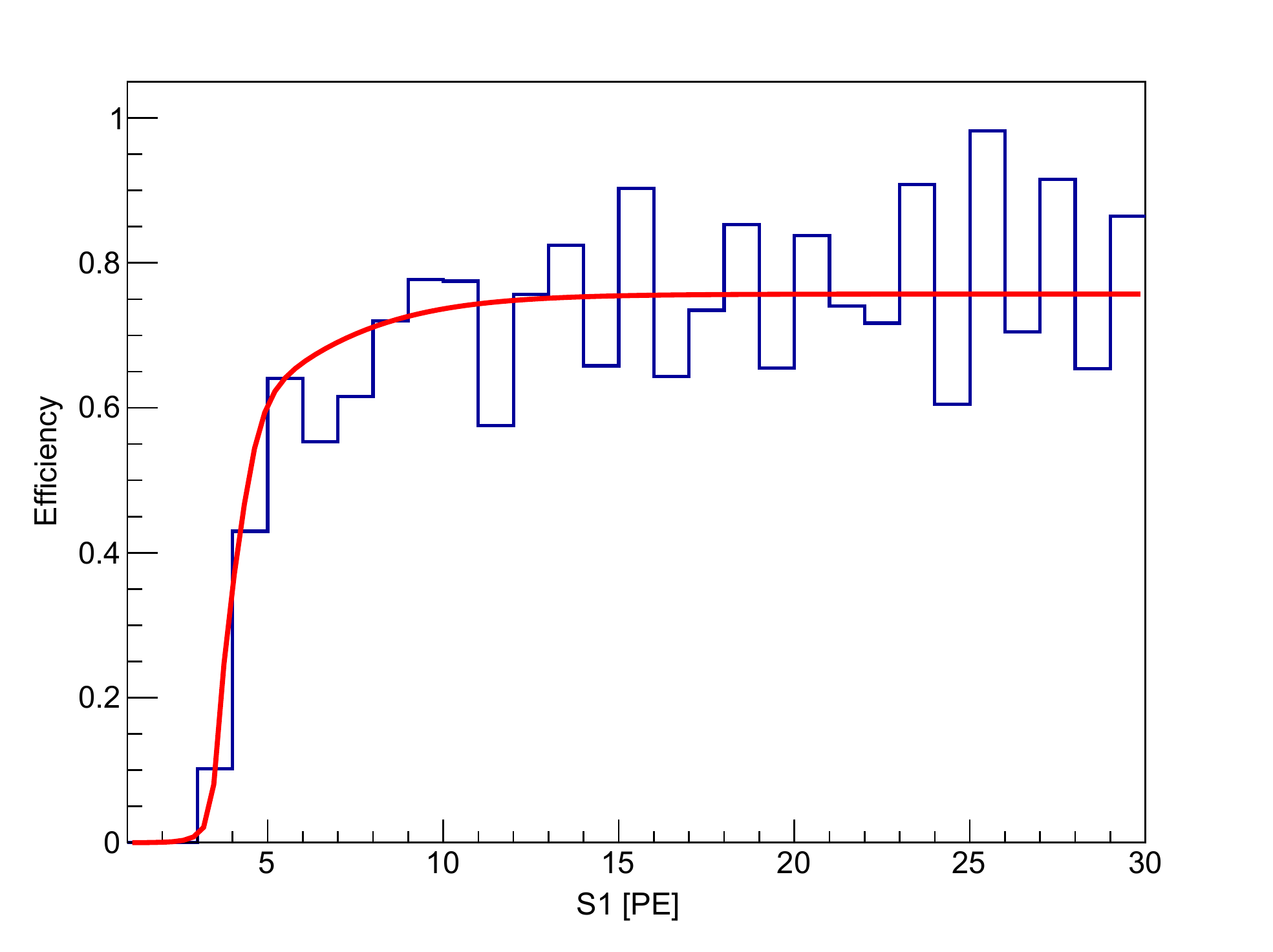}
  }
  \caption{
    (a) The band of $\log_{10}({\rm S2}/{\rm S1})$ versus S1 for the $^{60}$Co 
    ER calibration runs with
    the median and $\pm$2$\sigma$ of the band indicated as the solid and
    dashed blue lines, respectively.
    The median of the NR band is indicated as the solid red line, below which the 12
    leaked events are plotted as green markers.
    The dashed magenta line
    is the 300 PE cut on S2.
    The gray
    dashed lines are the equal energy lines with ER energy indicated 
    in the figures;
    (b) The ER efficiency in S1 obtained by taking the ratio 
    between the data and MC (histogram), and the red curve is a fit to the efficiency.
  }
\end{figure}

The ER calibration is performed with a $^{60}$Co gamma source,  interleaved frequently during dark
matter data taking. Low-energy $\gamma$ rays are produced through the well-known Compton scattering mechanism.
The distribution of the
single scatter ER events in log$_{10}({\rm S2}/{\rm S1})$
vs S1 is shown in Fig.~\ref{fig:ER}. All cuts, including a fiducial cut, have been applied.
For events with S1 between 2 and 30 PE, 12 out of 1,520 events were located 
below the median of NR in log$_{10}({\rm S2}/{\rm S1})$.
Subtracting the expected 1.65 events from accidental coincidence (see later discussions),
the remaining ER leakage is 0.68$\pm$0.23\% of the total, consistent with a pure
Gaussian expectation (0.5\%) obtained by fitting the ER band distribution.

Within the S1 range of 10 to 30 PE, the efficiency for ER detection and selection is
estimated to be 75.7\% by taking the ratio between the final number of events after all
cuts and the raw events on the ER band. At lower energy, the efficiency is estimated
by taking the ratio between the measured and expected S1 spectrum from MC
with 75.7\% at higher energy as an anchor (shown in Fig.~\ref{fig:ER_eff}). 
The overall efficiency is approximately 71.5\% in entire 2--30 PE range.



\section{Backgrounds in Dark Matter Search Data}

\label{sec:bkg}
The low-energy dark matter window was blinded in the analysis until all data
cuts were determined. The cuts on S1 and on the fiducial volume were optimized
from a figure-of-merit based on the expected below-NR-median
backgrounds of the ER,
the accidental background (statistically determined from data), and the
neutron background (MC estimates). The final optimized
search window on S1 is from 2 to 30 PE, and that on S2 is 300 to 10,000 PE. The
fiducial cut is determined as $r^2<500$\,cm$^2$ with a drift time
between 10 to 80\,$\mu$s, resulting in a fiducial mass of 54.0$\pm$2.3\,kg.
In what follows, we shall discuss the background
contributions in the dark matter search.

\paragraph{ER background}
Expected ER background in our final candidate sample with all cuts imposed,
summarized in Table~\ref{tab:bkg_rates},
has been estimated with a GEANT4-based MC program, with a few updates
compared to that in Ref.~\cite{pandax:first}.
\begin{table}[!htbp]
  \centering
  \begin{tabular}{lc}
    \hline
    Source & background level (mDRU) \\\hline
    Top PMT array & 4.7$\pm$2.3 \\
    Bottom PMT array & 2.3 $\pm$1.5 \\
    Inner vessel components & 3.8 $\pm$2.2 \\
    TPC components & 1.9 $\pm$0.9\\
    $^{85}$Kr & 2.6 $\pm$ 1.2\\
    $^{222}$Rn \& $^{220}$Rn & 0.5 $\pm$ 0.2\\
    Outer vessel & 0.9$\pm$0.6 \\\hline
    Total expected & 16.7$\pm$3.9\\\hline
    Total observed & 23.6$\pm$3.5\\\hline
  \end{tabular}
  \caption{The expected and observed background rates in the fiducial volume and
    in the dark matter search
    window. mDRU = 10$^{-3}$ evt/day/kg/keV$_{ee}$.
    Uncertainties in the MC prediction originate from uncertainties in the material
    radioactivity screening, except those for Rn and Kr which are due to the
    uncertainties in the PandaX data. }
  \label{tab:bkg_rates}
\end{table}
First, by taking into account the additional energy deposition in the
below-cathode region (``X'' events) observable through PMT arrays,
some of the MC events shifted above of the dark matter search
window, leading to a reduction of background from almost all components.
Second, the radioactivity level of the stainless steel vessel
was updated with a counting measurement
with much better statistics, also resulting
in a reduction in background expectation. Third, the internal
$^{85}$Kr, $^{222}$Rn and $^{220}$Rn levels were studied with the
statistics of the full dark matter search data sample with the same delayed coincidence techniques
as in Ref.~\cite{pandax:first}. The measured Kr concentration in Xe is
$68\pm29$ ppt mole/mole (uncertainties mainly due to event selection methods in the
analysis) assuming a
$2\times10^{-11}$ isotopic abundance of $^{85}$Kr,
leading to an expected background of
$2.6\pm1.2$ mDRU based on the MC.
The $^{222}$Rn and $^{220}$Rn backgrounds were determined to be
$0.7\pm0.2$ and $0.15\pm0.06$ mBq in the fiducial volume,
respectively, with uncertainties primarily arising from event selection cuts.
The resulting background of $0.5\pm0.2$ mDRU in the dark matter energy and fiducial volume search
window is estimated by MC
with an improved treatment taking into account the non-secular equilibrium
due to the long-lived isotope $^{210}$Pb ($\tau=22.2$ year).
The overall ER background in the dark matter search data estimated from radioactivity counting
is 16.7$\pm$3.9 mDRU.
This is consistent with the ER background 23.6 mDRU extrapolated from events with
S1$>30$ after efficiency correction
($\pm$15\% depending on the energy cut as well as efficiency modeling), 
assuming a flat distribution of the ER background in keV$_{ee}$ at
very low ER energy based on the MC.

The real relevant ER background for dark matter searches 
is formed from events that leak below
the NR-median, including those due to detector effects as well as the so-called
``gamma-X'' events with partial energy deposition in the ``chargeless'' regions.
To reliably estimate the number of such events, it is best
to use the ER calibration data where such events are
included with the right proportion.

\paragraph{Neutron background}
The neutron background is estimated using a combination of
SOURCES-4A~\cite{ref:source4a}
and GEANT4 simulation, leading to an estimate of
1.45 events within the 54.0$\times$80.1 kg-day
exposure before efficiency cuts, and about 0.35 events after all cuts.
This yields 0.18 neutron background events
below the NR medium line. We assign a generous 50\% uncertainty to the MC estimate.
Alternatively, a 90\%-confidence-level upper limit of 1.15 neutron
events can be set based on
the single to multiple NR scattering ratio from the MC and the absence of the
multiple scattering NR's in the dark matter search.

\paragraph{Accidental background}
In our dark matter search data, we find 
a significant number of isolated S1 and S2 events, which
yield a substantial background.
An isolated S1 is an event occurring without an obvious S2 nearby.
These signals are likely from multiple origins, e.g.
light leaking into the TPC due to interaction in the skin region, small discharges in the TPC
due to impurities or high voltage, and the accidental coincidence of SPE between PMTs.
An isolated S2 is an event without an S1 proceeding the waveform, which can be
due to events with very low energy of which S1 cannot be detected.
In addition, based on a visual inspection of isolated S2 events,  it was noticed that
a significant fraction of such events have a spiky timing
profile at the beginning of S2 (but cannot be efficiently rejected
with existing algorithms), implying that these S2 events happened very close to the gate grid where
S1 and S2 can no longer be separated.

\begin{figure}[!htbp]
\centering
\subfigure[Isolated S1]
{
  \label{fig:singles1}
  \includegraphics[width=0.45\textwidth]{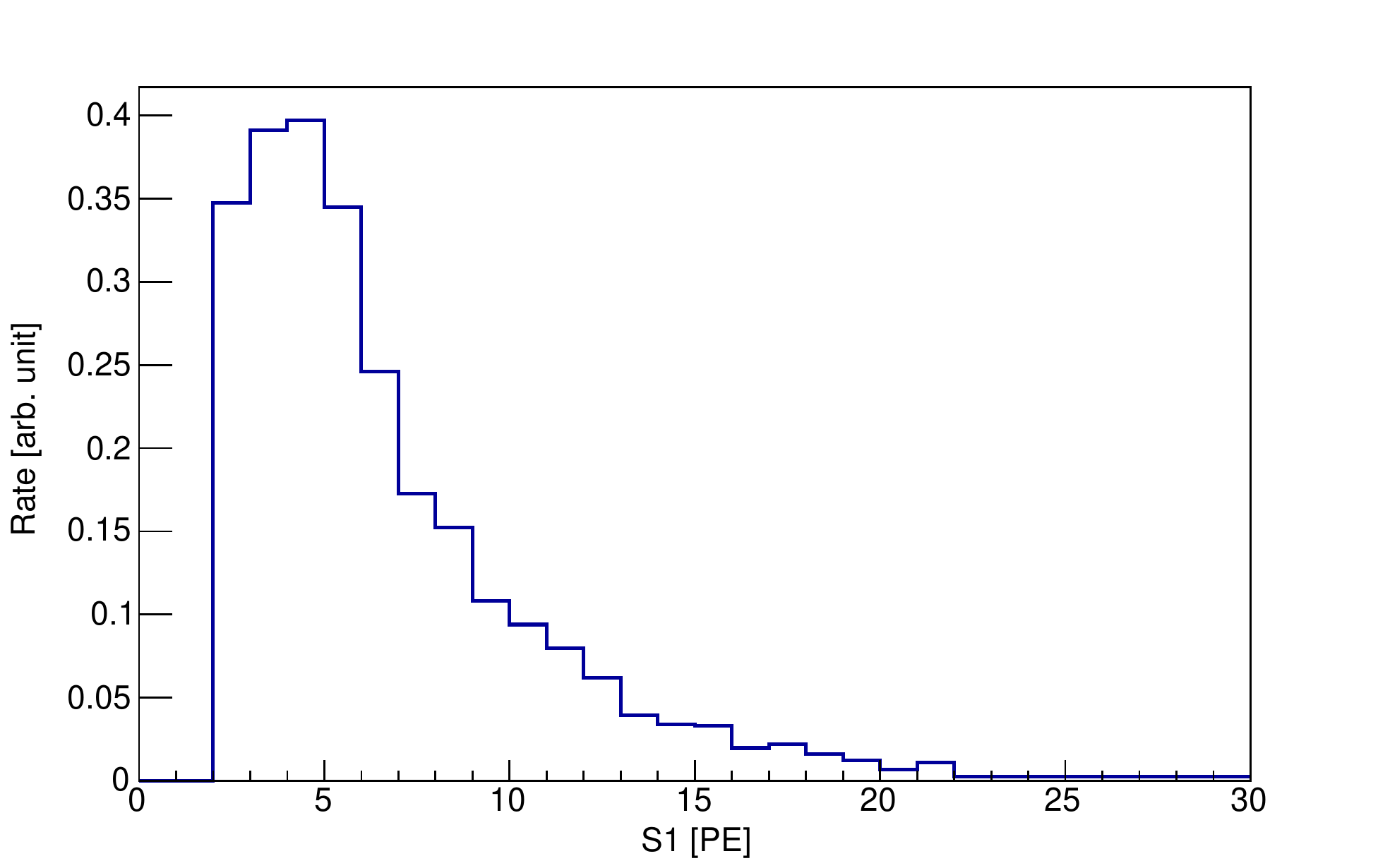}
}
\subfigure[Isolated S2]
{
  \label{fig:singles2}
  \includegraphics[width=0.45\textwidth]{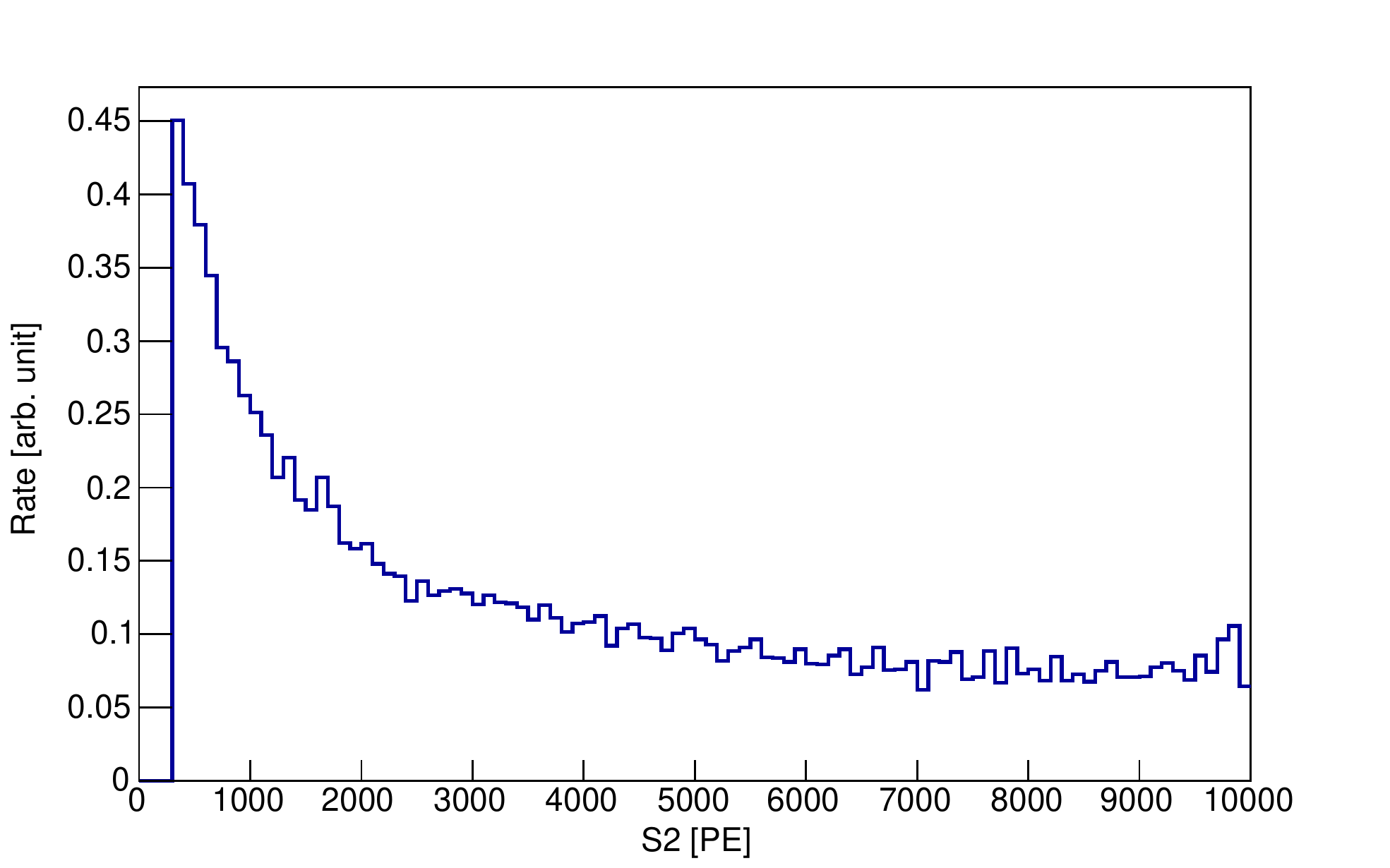}
}
\subfigure[Accidental]
{
  \label{fig:acc}
  \includegraphics[width=0.45\textwidth]{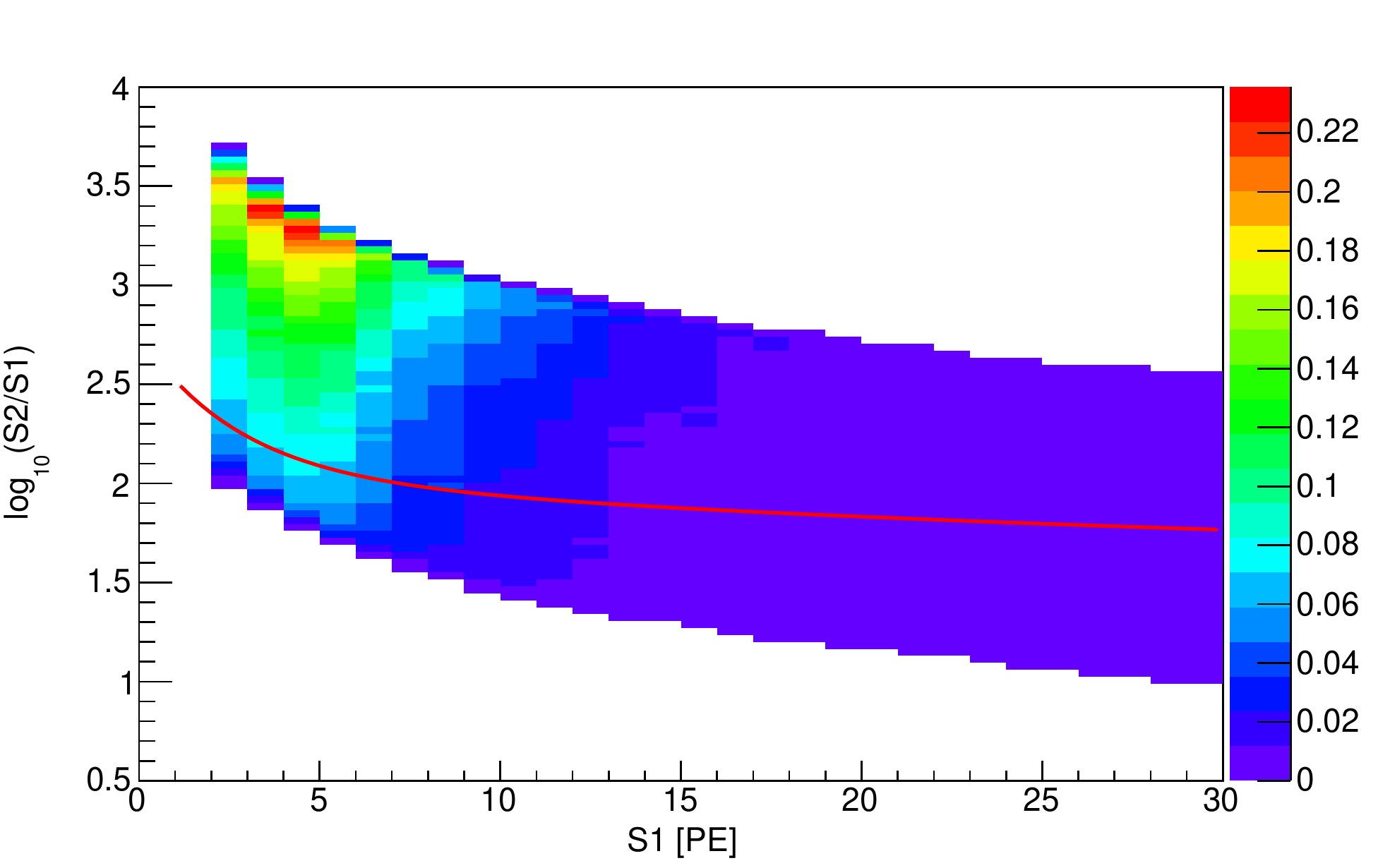}
}
\caption{
  The measured isolated S1 (a) and S2 (b) charge spectra, and the
  band of $\log_{10}({\rm S2}/{\rm S1})$ versus S1 (c) for the
  statistically obtained accidental background, with the sharp 
  cutoffs at the top and bottom corresponding to the 10,000 and 300 PE cut. 
  The median of the NR band is indicated as the red line. 
}
\end{figure}

In our dark matter
data, isolated S1 events are
estimated by looking for uncorrelated S1 events before a large S1 (which is associated with a trigger) yielding
a rate of about 23\,Hz, with the charge distribution shown
in Fig.~\ref{fig:singles1}. Isolated S2 events are measured with a rate of about 240 events/day for S2
within 300 to 10,000 PE (the 300 PE cut is imposed to balance between the
suppression of such background
and the loss of low-energy sensitivity), without obvious non-uniformity
in the horizontal plane. The charge spectrum of
such S2 signals is shown in Fig.~\ref{fig:singles2}.
The rate of such events
in the $^{60}$Co calibration runs increased to about 568 events/day,
and the amount of the rate increase is consistent with the MC expectation. 

Isolated S1 and S2 events produce accidental coincidences
which mimic real events. Such background can be statistically evaluated 
by forming random pairs in S1 and S2, and the resulting distribution in 
log$_{10}$(S2/S1) versus S1 is shown in Fig.~\ref{fig:acc}.
The overall rate in the dark matter data is estimated to be
35.1 events in 80.1 day with a conservative 10\% uncertainty based on the statistical 
uncertainty of a day-long dark matter search run.

\section{Candidate Events From 80.1 Day Dark Matter Search Data}

\begin{figure}[!htbp]
\centering
\subfigure[drift time vs. $r^2$]
{
  \label{fig:dm_vert}
  \includegraphics[width=0.45\textwidth]{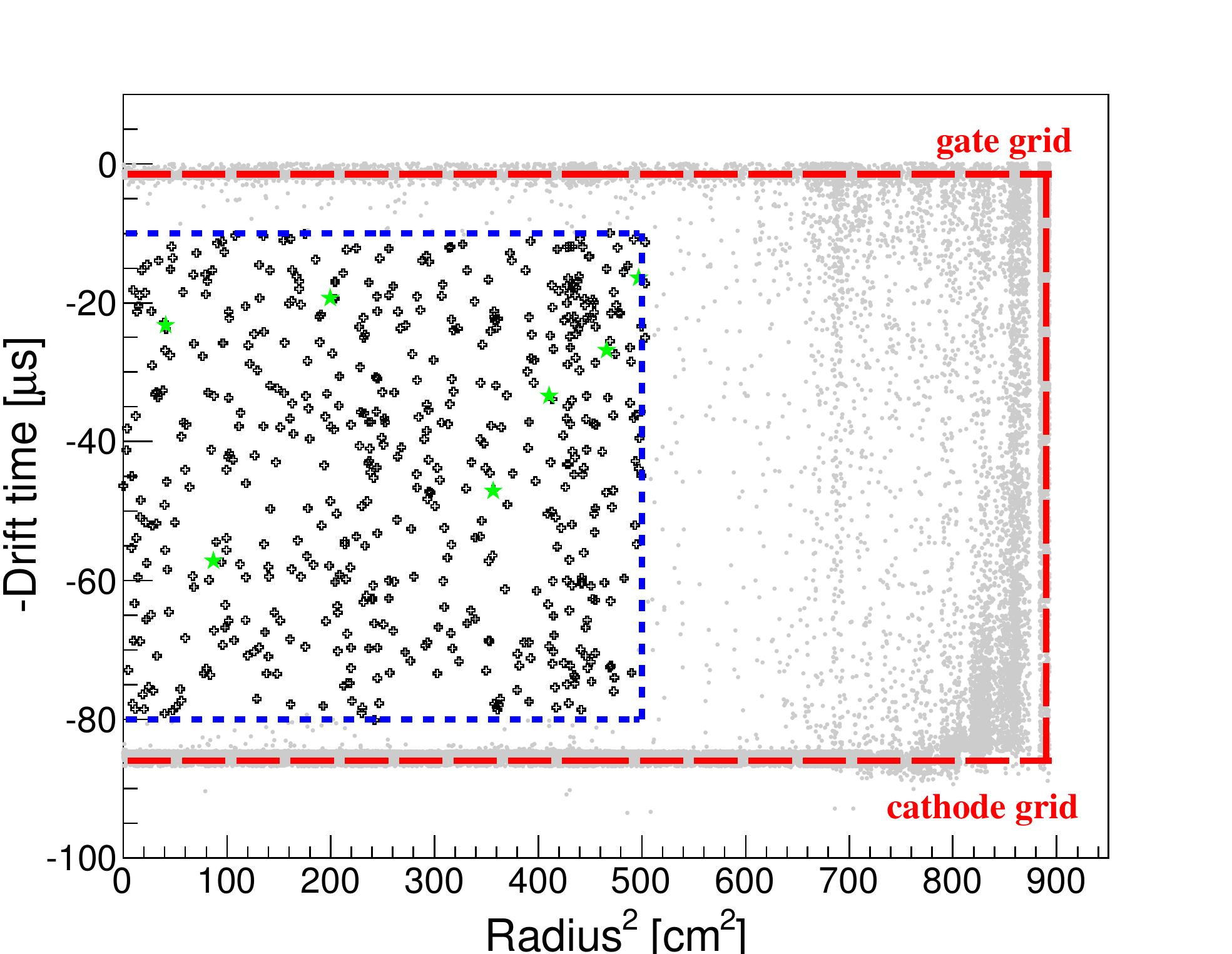}
}
\subfigure[$r^{2}$]
{
  \label{fig:r2_check}
  \includegraphics[width=0.45\textwidth]{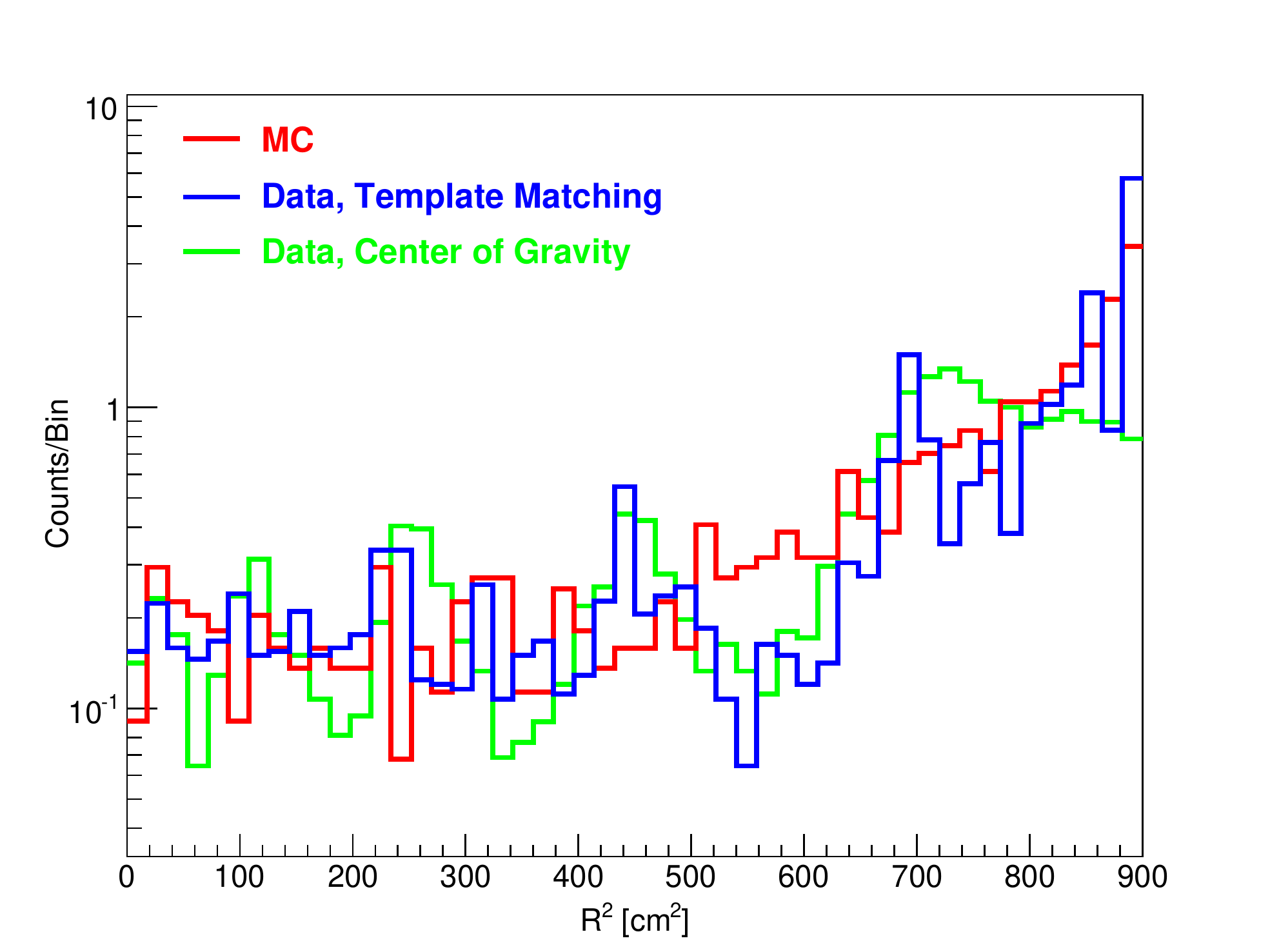}
}
\subfigure[drift time]
{
  \label{fig:z_check}
  \includegraphics[width=0.45\textwidth]{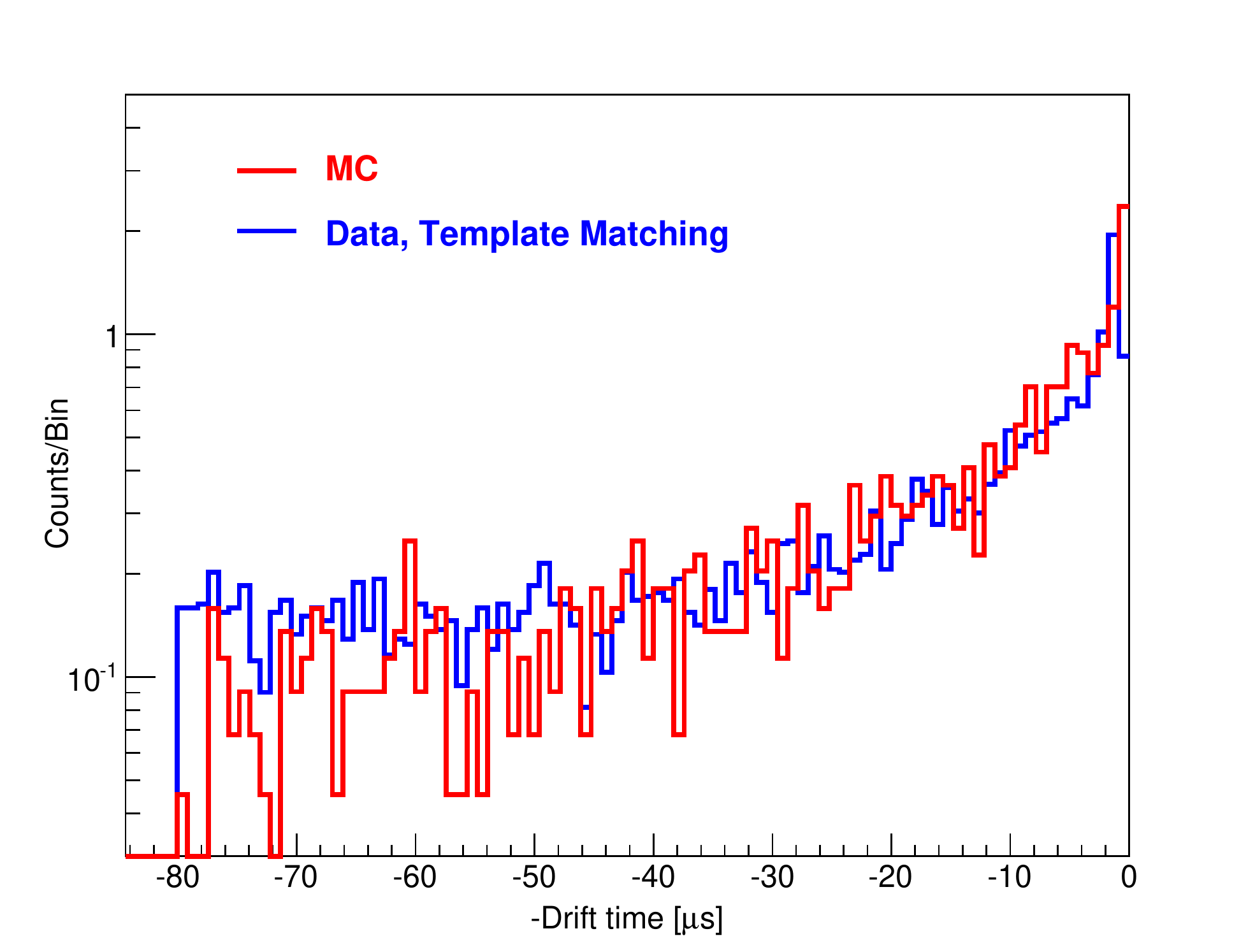}
}
\caption{
(a) The vertex distribution of events in the TPC during dark matter data taking. The
blue dashed box indicates the fiducial volume, and the red dashed box indicates the
entire active volume within the TPC, confined by the cathode, gate grid, and the PTFE wall.
The events below the NR median are indicated by the green markers;
(b) projection in $r^2$, data (TM method for position reconstruction) in blue, data (CoG method) in green, and the MC energy deposition CoG in red;
(c) projection in the drift time, data (TM) in blue and the MC energy deposition CoG in red.
}
\label{fig:DM_band_vt}
\end{figure}

For the dark matter search data, the event rates after
different levels of cuts are summarized in Table~\ref{tab:DM_evt_list}.
The data quality cuts remove a large fraction of the
multiple scattering events, reducing the total number considerably,
which also explains that the subsequent single-site cut
has a small effect on the remaining number of events.
\begin{table}[!htbp]
\renewcommand{\tabcolsep}{0.25 cm}
\centering
\begin{tabular}{lrc}
 \hline
 Cut             &  \# Events  &  Rate (Hz) \\
 \hline
 All triggers    &  24,762,972  &  3.58     \\
 Quality cut     &  6,127,280   &  0.88     \\
 Single-site cut &  5,050,845   &  0.73    \\
 S1 range (2--30 PE) & 62,872    &  $9.08\times10^{-3}$ \\
 S2 range (300--10,000 PE) & 44,171 & $6.38\times10^{-3}$ \\
 Fiducial volume &  542      &  $7.83\times10^{-5}$ \\
 \hline
\end{tabular}
\caption{The event rate in the DM run after various level of analysis cuts.}
\label{tab:DM_evt_list}
\end{table}
Within prescribed cuts, 542 events were found in 54.0 kg $\times$ 80.1 days.
The event distribution in $r^2$ vs. drift time in the TPC
is shown in Fig.~\ref{fig:dm_vert}.
The event projections in $r^2$ (with two position reconstruction methods)
and drift time are also compared to the expected ER
distribution from the Monte Carlo, where good agreement is achieved.

\begin{figure}[!htbp]
\centering
\includegraphics[width=0.49\textwidth]{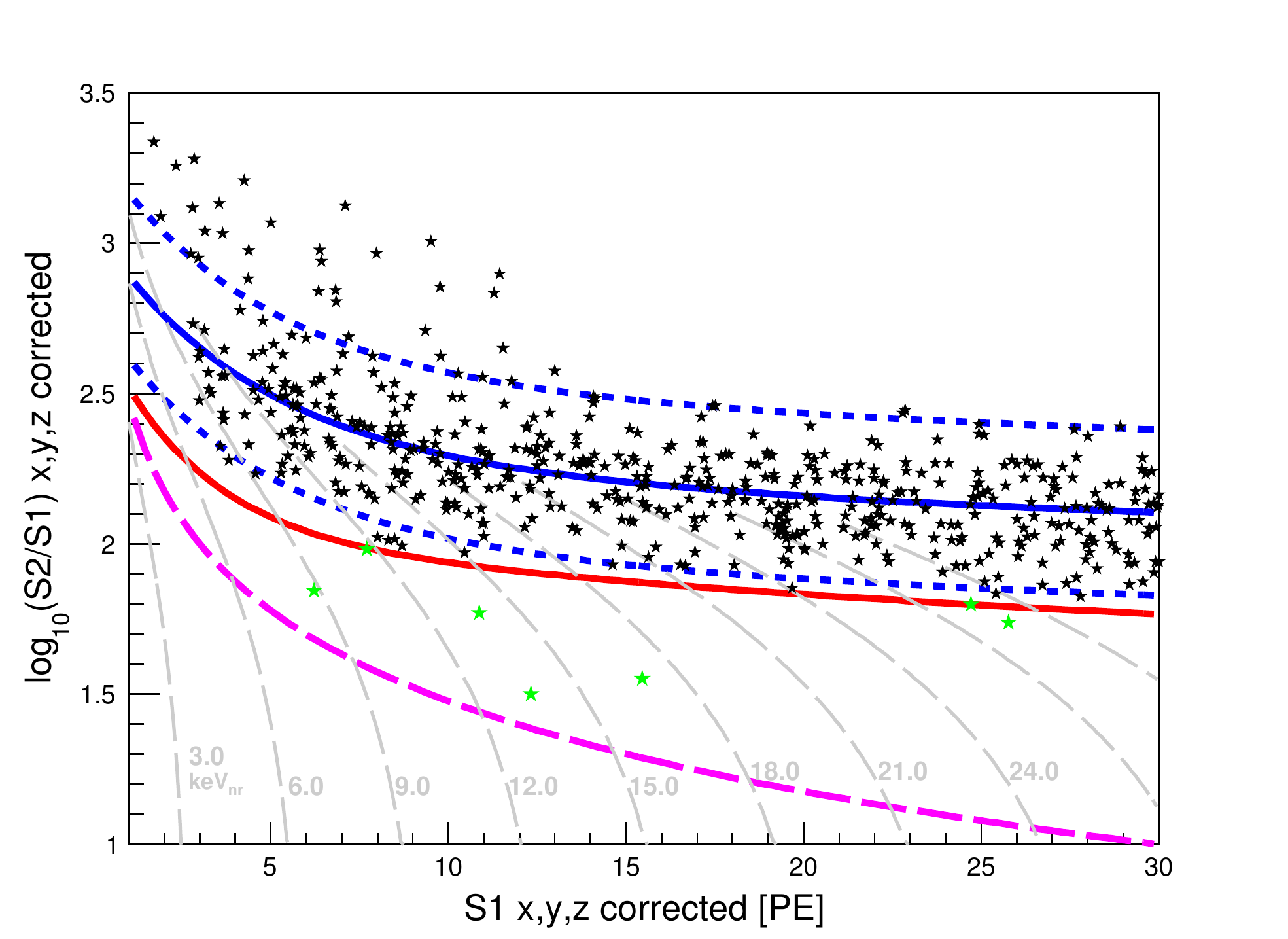}
\caption{The band of $\log_{10}({\rm S2}/{\rm S1})$ versus S1 for the dark matter search data. The ER
band is indicated by the blue solid line (median) and the dashed line ($\pm$2$\sigma$).
The median of the NR band is indicated as the solid red line. The dashed magenta line
is the 300 PE cut on S2. The green stars represent events below the NR median.
The gray
dashed lines are the equal energy lines with NR energy indicated 
in the figures.
}
\label{fig:dm_band}
\end{figure}

The distribution of events in log$_{10}({\rm S2}/{\rm S1})$ vs. S1 is shown
in Fig.~\ref{fig:dm_band}. The majority of the events are consistent with an
ER origin. The events located higher than the ER band at low S1 are
the accidental backgrounds, more prominent than those in the $^{60}$Co calibration run
due to the much lower ER event rate in the dark matter search data.
Seven of the candidate events are located below the median of the NR band
indicated by the green markers in Figs.~\ref{fig:dm_vert} and \ref{fig:dm_band}.
For comparison, the expected background in the total sample as well as those
below the NR mean is shown in Table~\ref{tab:dm_rates}. The ER background
is estimated based on the 23.6 mDRU value, a corresponding ER
energy range of 6.9\,keV$_{ee}$, and an average ER efficiency of 71.5\%.
The below-NR-median accidental background is estimated based on the distribution in
Fig.~\ref{fig:acc}.
Summing over all the contributions, we expect 6.9 background events
below the NR median.  No significant excess above the background is observed.
\begin{table}[!htbp]
\renewcommand{\tabcolsep}{0.1cm}
  \centering
  \begin{tabular}{lccccc}
    \hline
    & \multirow{2}{*}{ER} & \multirow{2}{*}{Accidental} & \multirow{2}{*}{Neutron} & Total & Total\\
    &    &            &         & expected & observed\\ \hline
    \multirow{2}{*}{All} & \multirow{2}{*}{503.7} & \multirow{2}{*}{35.1} & \multirow{2}{*}{0.35} & \multirow{2}{*}{539.1} & \multirow{2}{*}{542}\\
    & & & & & \\\hline
    Below & \multirow{2}{*}{2.5} & \multirow{2}{*}{4.2} & \multirow{2}{*}{0.18} & \multirow{2}{*}{6.9} &  \multirow{2}{*}7\\
    NR med & & & & &  \\\hline
  \end{tabular}
  \caption{The expected and measured events (in units of events) in 80.1\;live-day dark matter search data.}
  \label{tab:dm_rates}
\end{table}

\section{Fitting method}
To maximally use the information from the data,
instead of choosing only the below-NR-median
region to search for DM like in Ref.~\cite{pandax:first}, in this analysis we
defined a much extended DM window with S1 between 2 and 30 PE and S2 between 300 to
10,000 PE. To fit all data, an unbinned extended likelihood function is constructed as
\begin{equation}
\begin{aligned}
\label{eq:likelihood}
\mathcal{L} &= \rm{Poisson}(N_m|N_{exp})\times\\
 &\displaystyle{\Pi_{i=1}^{i=N_m}}[\frac{N_{DM}(1+\delta_{DM})P_{DM}({\rm S1}^i,{\rm S2}^i)\epsilon_{NR}({\rm S1}^i,{\rm S2}^i)}{N_{exp}}\\
 &+ \frac{N_{ER}(1+\delta_{ER})P_{ER}({\rm S1}^i,{\rm S2}^i)}{N_{exp}} \\
 &+ \frac{N_{Acc}(1+\delta_{Acc})P_{Acc}({\rm S1}^i,{\rm S2}^i)}{N_{exp}} \\
 &+ \frac{N_{nbkg}(1+\delta_{nbkg})P_{nbkg}({\rm S1}^i,{\rm S2}^i)\epsilon_{NR}({\rm S1}^i,{\rm S2}^i)}{N_{exp}}]\\
 &\times G(\delta_{DM},0.2)G(\delta_{ER},0.15)G(\delta_{Acc},0.1)G(\delta_{nbkg},0.5) \, ,
\end{aligned}
\end{equation}
where $N_m$ and $N_{exp}$ are the total number of measured and fitted candidates with
\begin{equation}
\begin{aligned}
\label{eq:total_evt}
  N_{exp} &= N_{DM}\langle\epsilon_{NR}\rangle_{DM}(1+\delta_{DM}) + N_{ER}(1+\delta_{ER}) \\
 &+ N_{Acc}(1+\delta_{Acc}) + N_{nbkg}\langle\epsilon_{NR}\rangle_{nbkg}(1+\delta_{nbkg})\,.
\end{aligned}
\end{equation}
As indicated in Fig.~\ref{fig:DM_band_vt}, the position dependence of events in the 
fiducial volume is rather weak and is therefore ignored here for simplicity.
$N_{DM}$ ($N_{nbkg}$) is the total number of WIMP particles (neutrons) interacting with
the detector during the measurement before efficiency and acceptance cuts.
$N_{DM}$ is computed for each given pair of WIMP mass and cross section
$(m_{\chi}, \sigma_{n-\chi})$ assuming the
isothermal DM halo model~\cite{Smith:2007,Savage:2006} with a
local dark matter density of 0.3\,GeV/$c^2$/cm$^3$, a circular velocity of
220\,km/s, a galactic escape velocity of 544 km/s, and an average earth
velocity of 245\,km/s. $P_{DM}({\rm S1}^i,{\rm S2}^i)$ and $P_{nbkg}({\rm S1}^i,{\rm S2}^i)$
are the probability distribution functions (PDFs)
of NR recoil signals for a WIMP with given mass
and neutron background, respectively, obtained using the NEST-based
MC simulation employing the PDE, EEE, and SEG described earlier. $\epsilon_{NR}({\rm S1}^i,{\rm S2}^i)$
is the NR detection efficiency from Fig.~\ref{fig:NR_eff}, with acceptance
set to zero if S1 and S2 are outside the ranges of
$(2,30)$ and $(300, 10,000)$ PEs.
To obtain the expected measured dark matter and
neutron background events, the NR efficiency function has to be averaged over the
expected dark matter or neutron background PDF ($\langle\epsilon_{NR}\rangle_{DM}$ and
 $\langle\epsilon_{NR}\rangle_{nbkg}$) in Eq.~\ref{eq:total_evt}).
$N_{ER}$ and $N_{Acc}$ are the total number of ER and accidental background with
detection efficiency taken into account,
and $P_{ER}({\rm S1}^i,{\rm S2}^i)$ (taken to be the same as that obtained from ER calibration
from Fig.~\ref{fig:ER}, supported by the MC)
and $P_{Acc}({\rm S1}^i,{\rm S2}^i)$ (from Fig.~\ref{fig:acc}) are the
corresponding PDFs.
The contamination of the accidental background in the ER calibration
run is neglected due to the dominating ER rate in the calibration runs.
The expected background events are taken from the top row of
Table~\ref{tab:dm_rates}. To allow systematic variation in the global efficiency,
four normalization nuisance parameters ($\delta_{DM}$, $\delta_{ER}$, $\delta_{Acc}$
and $\delta_{nbkg}$) are included for the four type of events, constrained by
Gaussian variations ($G$'s in Eq.~\ref{eq:likelihood})
of 20\% (DM), 15\% (ER), 10\% (accidental) and 50\% (neutron background)
in the penalty terms~\cite{ref:cowan, ref:xenon_ll}.

\begin{figure}[!htbp]
  \centering
  \includegraphics[width=0.49\textwidth]{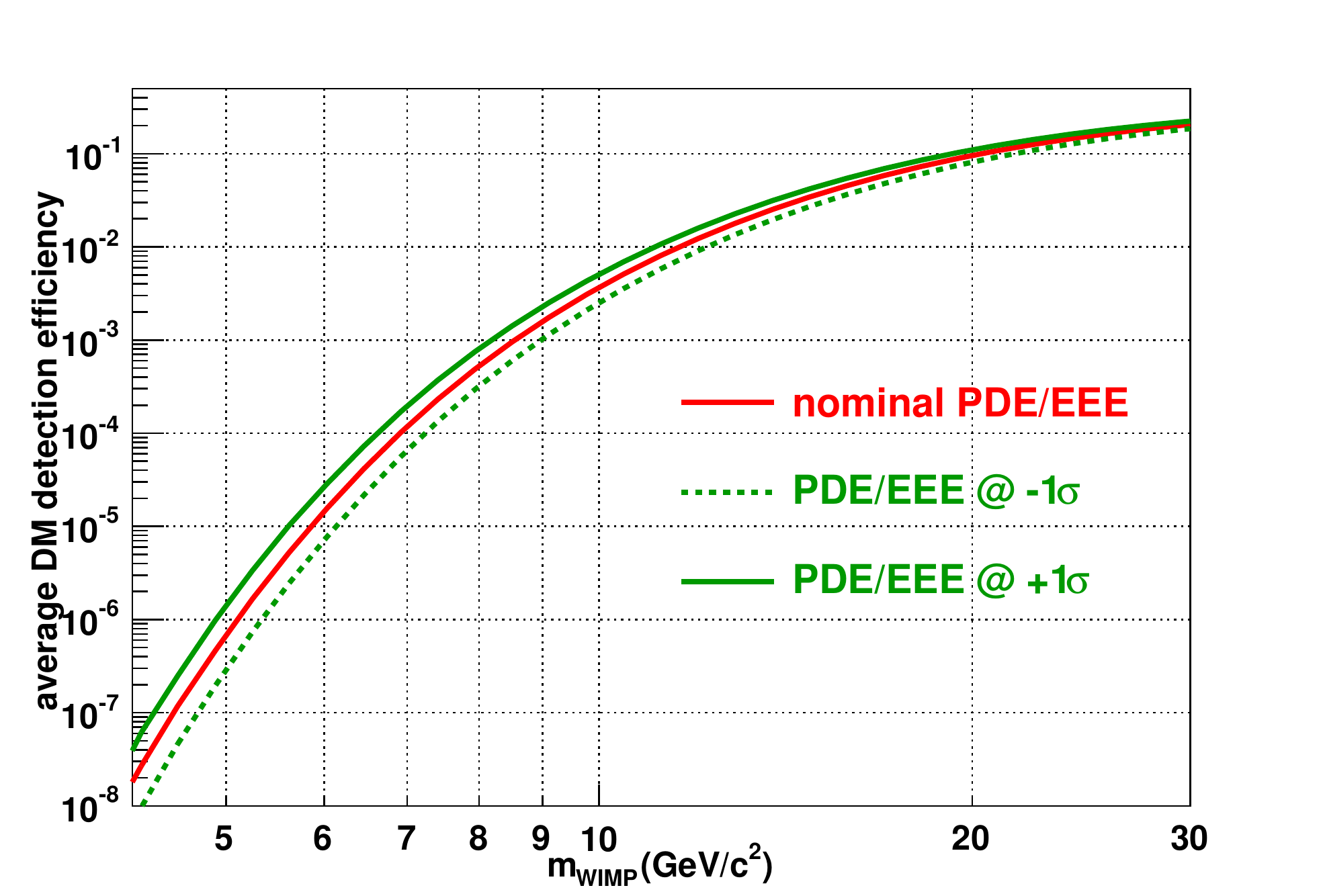}
  \caption{The overall detection efficiency for WIMP at different mass (red) and those
  obtained with PDE and EEE set at $+1\sigma$ (solid green) 
  and $-1\sigma$ (dashed green).}
  \label{fig:dm_eff}
\end{figure}

\begin{figure}[!htbp]
  \centering
  \includegraphics[width=0.49\textwidth]{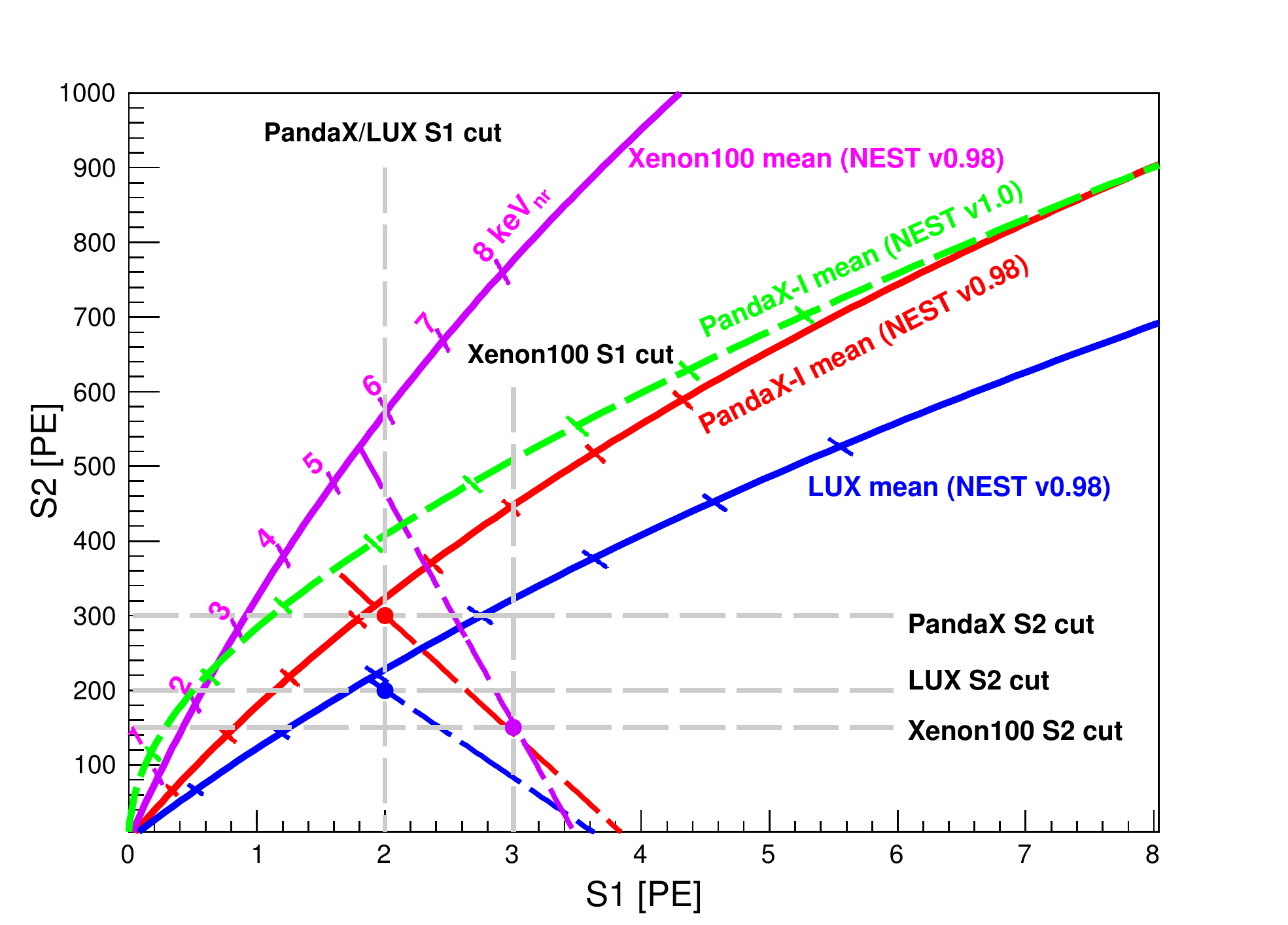}
  \caption{The comparison of the mean energy thresholds, translated 
    from the cuts on S1 and S2, for different experiments under the same energy 
    model (NEST v0.98). The solid curves (red: PandaX, violet: XENON100, blue: LUX) 
    represent the mean values 
    for S1 and S2 obtained from NEST with slanted ticks (along the equal-energy vector) 
    indicating the corresponding mean NR energy in divisions of
    keV$_{nr}$. The curve for PandaX based on NEXT v1.0 is drawn as the 
    green dashed curve. 
    The selection thresholds in S1 and S2 are indicated in the figure as the solid circles.
    The anti-diagonal dashed lines (red: PandaX, violet: XENON100, blue: LUX) 
    are the equal-energy lines projected from the corresponding threshold 
    points for different experiments.}
  \label{fig:dm_cut}
\end{figure}

The average WIMP detection efficiency $\langle\epsilon_{NR}\rangle_{DM}$,
obtained by combining the NR efficiency with the WIMP PDF (with fluctuations in S1 and S2 
properly taken into account), strongly
depends on the WIMP mass, which is depicted in Fig.~\ref{fig:dm_eff}. 
The lower the WIMP mass, the softer the recoil energy distribution, therefore
the selection threshold on S1 and S2 would more strongly 
suppress the overall efficiency. To compare the effects of selection thresholds of 
different experiments on the NR energy, the mean NR energy curves 
in S1 and S2 are plotted
in Fig.~\ref{fig:dm_cut} based on the NEST-0.98 model with the PDE, EEE, and
SEG values from PandaX, XENON100~\cite{ref:xenon100_detailed}, 
and LUX~\cite{ref:lux}. 
Our selection threshold is at about 4.2\,keV$_{nr}$
in both S1 and S2.
XENON100 achieved a S2 threshold of less than 2 keV$_{nr}$ but 
a much higher S1 threshold of about 8 keV$_{nr}$. 
LUX, on the other hand, achieves an average 3\,keV$_{nr}$
threshold on both S1 and S2, but in Ref.~\cite{ref:lux} they 
choose to drop the NR efficiency entirely below 3\,keV$_{nr}$.
The NEST-1.0 model
predicts a higher charge yield for NR, in which case our S1 threshold would stay, but
the S2 threshold would improve to about 2.8\,keV$_{nr}$, leading to a better sensitivity
for low mass WIMPs. Nevertheless, we chose the NEST-0.98 model to
report our final WIMP results.

\begin{figure}[!htbp]
  \centering
  \includegraphics[width=0.49\textwidth]{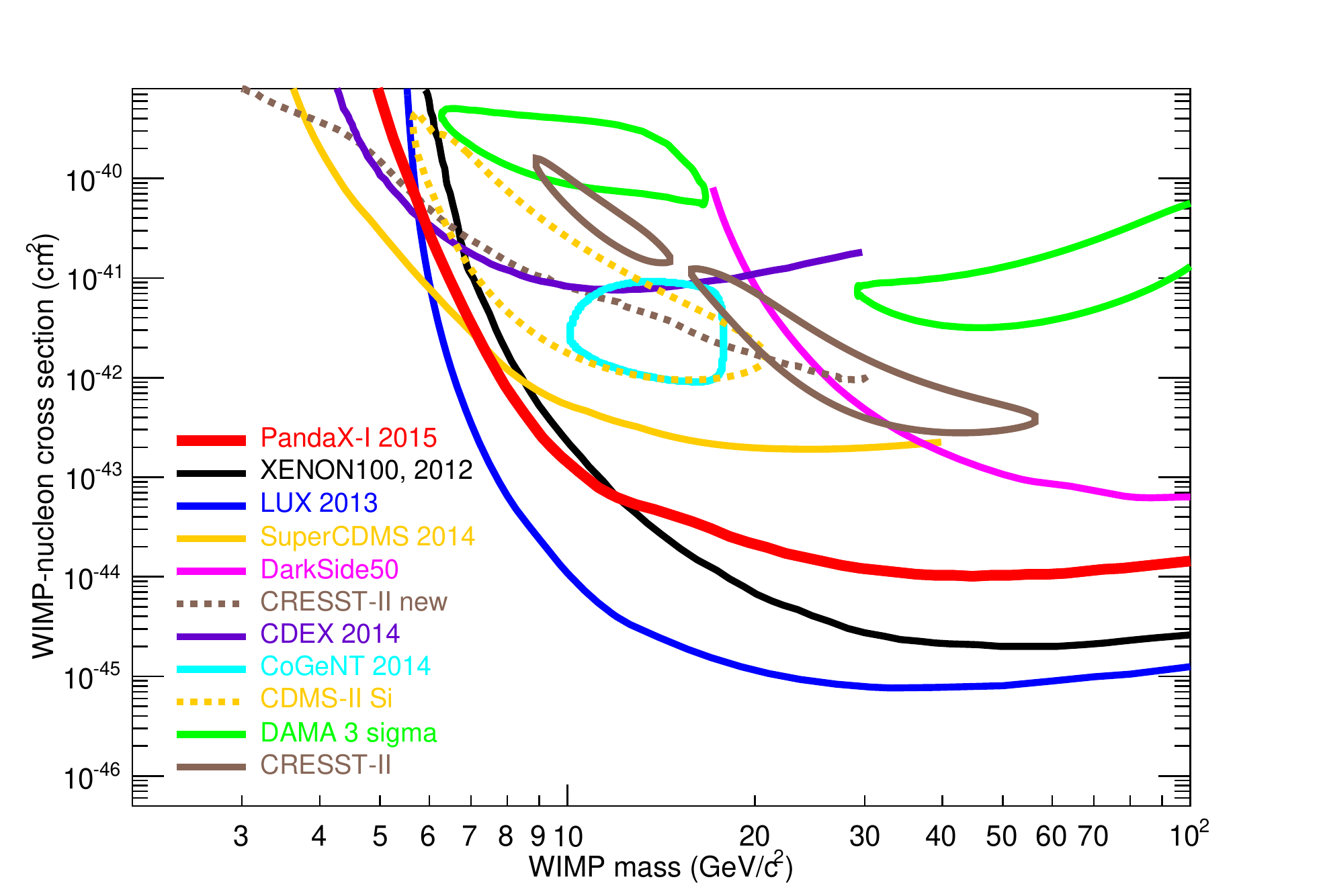}
  \caption{
    The 90\% c.l. upper limit for spin-independent isoscalar WIMP-nucleon cross
    section for the PandaX-I experiment (red curves).
    Recent world results are plotted for comparison: XENON100
    225 day results~\cite{xenon100_final} (black solid),  LUX first
    results~\cite{ref:lux} (blue),  SuperCDMS results~\cite{ref:supercdms} (orange
    solid), DarkSide results~\cite{ref:darkside} (magenta solid), CRESST-II 2014
    limits~\cite{ref:cresst2} (brown dashed), and CDEX 2014 limits~\cite{ref:cdex} (solid violet).
    The claimed WIMP signals are
    shown as closed contours: CoGeNT 2014 results~\cite{ref:cogent} (cyan solid),
    CDMS-II-Si results~\cite{ref:cdms-si} (gold dashed), DAMA/LIBRA 3$\sigma$
    contours~\cite{ref:dama_savage} (green solid), and CRESST-II 2012 results~\cite{ref:cresst1} (brown solid).
  }
  \label{fig:final}
\end{figure}

The best fit value to maximize the likelihood function is found at
$m_{\chi} = 27.5$~GeV/c$^2$ with a $\sigma_{\chi,N}$ at
4.1$\times10^{-45}$~cm$^2$. The value of the likelihood is also consistent with
that from the null hypothesis within $1\sigma$,
indicating no significant excess over the background.  To set the WIMP search
upper limit, a standard profile likelihood ratio statistic is formed~\cite{ref:cowan, ref:xenon_ll}. 
A Feldman and Cousin
approach~\cite{ref:fc} is used to fit the data as well as a large number of MC simulations
using the signal hypothesis at each grid point of $(m_{\chi}, \sigma_{n-\chi})$.
The 90\% c.l. upper limit obtained with this approach
is shown in Fig.~\ref{fig:final} together with the world data,
and is verified to be very similar to that
obtained assuming an approximate half-$\chi^2$
distribution of the test statistic~\cite{ref:cowan}.  A binned likelihood method
developed in the independent analysis yields an upper limit in good agreement with
the above.
The upper limit excludes a WIMP mass of 10\,GeV/c$^2$ down to a cross section of 1.41$\times10^{-43}$\,cm$^2$, and
the lowest excluded cross section is 1.01$\times10^{-44}$\,cm$^2$ at
a WIMP mass of 44.7\,GeV/c$^2$.
Under the elastic, spin-independent, and isospin conserving WIMP-nucleon scattering
model,  our limits strongly disfavor the WIMP interpretation of the results from
DAMA/LIBRA, CoGeNT, CDMS-II-Si and CRESST-II.
It is noteworthy that the PDE and EEE used in this analysis are 
conservative in nature since we inhibited the unstable PMTs. In addition, 
we have considered the average
WIMP detection efficiency with WIMP mass dependence in this analysis 
(Fig.~\ref{fig:dm_eff}).
Compared to that in Ref.~\cite{pandax:first}, 
in which the DM efficiency is treated only as a function of S1, this
treatment is more realistic. Even with these realistic treatments, our results still set a
stringent limit at the low WIMP mass region,
with a tighter
bound than SuperCDMS above the WIMP mass of 7\,GeV/$c^2$, and
the best reported bound in a dual phase xenon detector below a WIMP mass
of 5.5\,GeV/c$^2$. 
Note that one of the key difference between this analysis and that
from LUX in Ref.~\cite{ref:lux} 
is that the latter made a conservative choice to model no signal
generation for events below 3 keV$_{nr}$, while in our treatment the signals
generation is continuous to zero energy therefore low energy events below the mean 
energy threshold of 4.2 keV$_{nr}$ could still fluctuate upwards into
the detection region.

\begin{figure}[!htbp]
  \centering
  \includegraphics[width=0.49\textwidth]{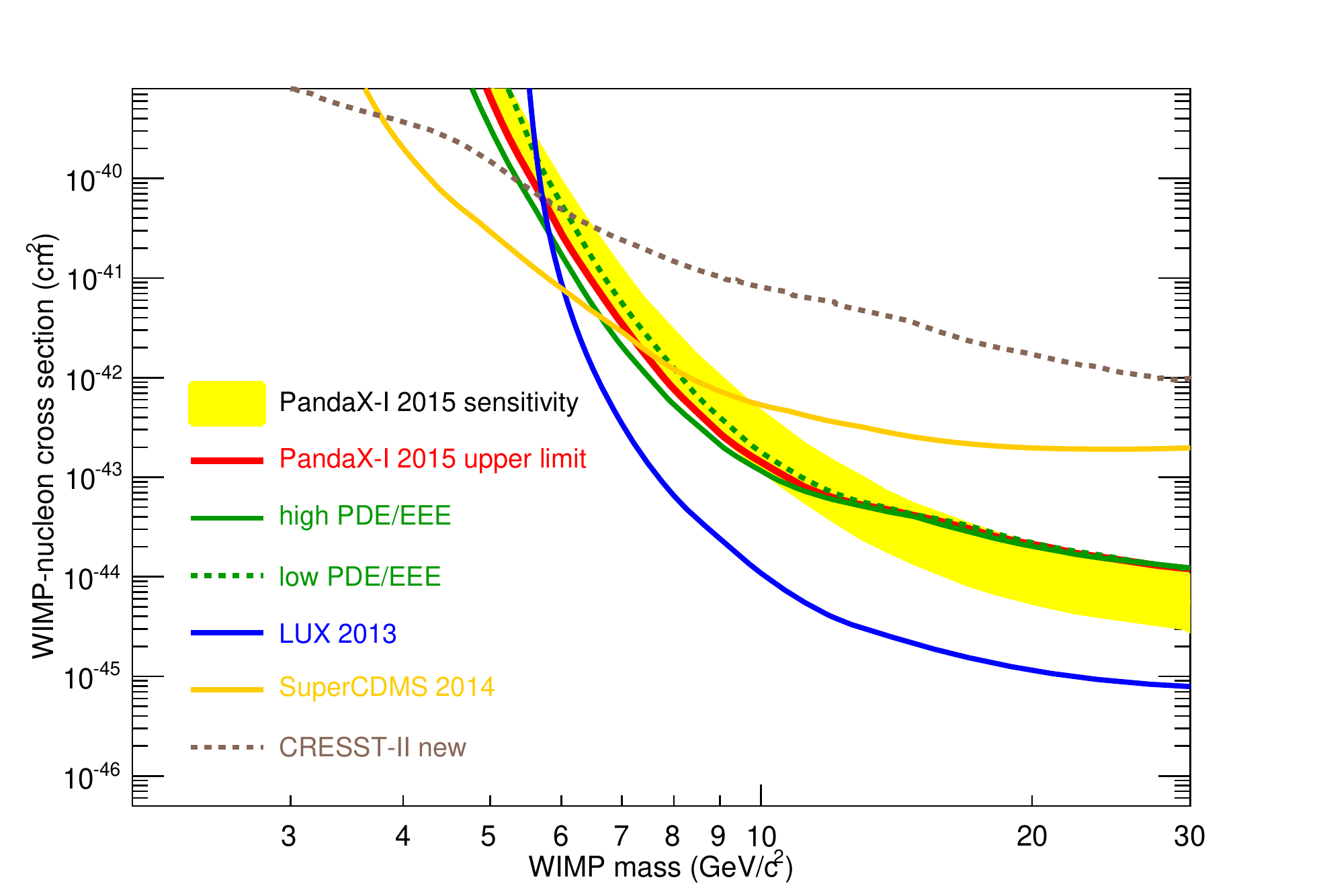}
  \caption{
    PandaX-I WIMP search limit from the data (red line)
    overlaid with the $\pm1\sigma$ sensitivity
    band obtained from toy MC (yellow) as well as the alternative upper limits
    using either $+1\sigma$ or $-1\sigma$ values for the PDE and EEE, but
    with the same NEST-0.98 model. For comparison,
    a few world leading limits for the low mass WIMP
    are plotted: LUX first
    results~\cite{ref:lux} (blue),  SuperCDMS results~\cite{ref:supercdms} (orange), and CRESST-II 2014
    limits~\cite{ref:cresst2} (brown dashed).
  }
  \label{fig:lim_sys}
\end{figure}

The experimental sensitivity band is obtained using the same approach as above but
with hundreds of 80.1-day 
background-only toy MCs based on Table~\ref{tab:dm_rates} 
using prescribed PDF for each event type, from which one obtains
a distribution of ``upper limits''.
In Fig.~\ref{fig:lim_sys}, our upper limit is overlaid with
the $\pm$1-$\sigma$ sensitivity band. Consistency is observed, confirming
no significant excess over background.

To study shape related systematic uncertainties separately~\footnote{The shape systematics could also be introduced into the fitter via nuisance parameters. However, to explicitly show the size of the effects and to simplify the fitter computation, we chose to apply these systematic variations ``by hand''.}, we performed calculations
of upper limits either by setting PDE and EEE both at $+1\sigma$ 
or $-1\sigma$. The resulting
limits are overlaid in Fig.~\ref{fig:lim_sys}. As expected, the higher efficiency would lead
to tighter bounds in the low mass region and vice versa.  The (more aggressive) upper limit
obtained with dark matter PDFs generated from the NEST-1.0 model is very
close to that with the $+1\sigma$ PDE/EEE. These are sizable influences but 
are comparable with the sensitivity band, therefore do not
change the main conclusion of our results.

\section{Conclusion and outlook}
In summary, we report the low-energy dark matter search results with the
54.0$\times$80.1 kg-day full exposure of the PandaX-I
experiment. In this analysis, compared to the first results,
we made a number of improvements in
signal identification,  background classification and rate and shape
estimates, a realistic treatment on the
efficiency for very low recoil energy events, as well as
profile likelihood ratio fits to obtain the final WIMP search limit.
Observing no significant excess over background, our
results strongly disfavor the WIMP interpretation of the results from
DAMA/LIBRA, CoGeNT, CDMS-II-Si and CRESST-II. Our bound is
tighter than that from SuperCDMS above the WIMP mass of 7\,GeV/$c^2$, and
is the lowest reported limit below a WIMP mass of 5.5\,GeV/$c^2$ in xenon dark matter
experiments to date, showing that liquid xenon detectors
can be competitive for low-mass WIMP searches.

The results from PandaX-I are crucial in guiding the future development
of the PandaX program.
The second phase experiment, PandaX-II, constructed with a liquid xenon target of 500 kg sensitive mass and lower background materials for the cryostat and TPC, is under
preparation at CJPL. The PandaX-II detector is expected
to improve both on the light and charge collection efficiency and
push the dark matter sensitivity beyond the current best reach in a wide range of
WIMP masses.

\section{Acknowledgement}
This project has been supported by a 985-III grant from Shanghai Jiao Tong
University, a
973 grant from Ministry of Science and Technology of China (No. 2010CB833005),
grants 
from National Science Foundation of China (Nos. 11055003 and 11435008), 
and a grant from the Office
of Science and Technology in Shanghai Municipal Government (No. 11DZ2260700).
This work is supported in part by the CAS Center for Excellence in Particle Physics (CCEPP).
Xun Chen acknowledges support from China Postdoctoral Science Foundation
Grant 2014M551395.
The project has also been sponsored by Shandong University, Peking University,
the University
of Maryland, and the University of Michigan. We would like to thank many people
including
Elena Aprile, Xianfeng Chen, Carter Hall, T. D. Lee, Zhongqin Lin, Chuan Liu, Lv Lv,
Yinghong Peng, Weilian Tong, Hanguo Wang, James White, Yueliang Wu,
Qinghao Ye,
Qian Yue, and Haiying Zhao for help and discussion at
various levels. We are particularly indebted to Jie Zhang for his strong support
and crucial
help during many stages of this project. Finally, we thank the following
organizations and
personnel for indispensable logistics and other supports: the CJPL administration
including
directors Jianping Cheng and Kejun Kang and manager Jianmin Li, Yalong
River Hydropower Development Company Ltd.  
including the chairman of the board Huisheng Wang, and manager
Xiantao Chen and his JinPing tunnel management team from
the 21st Bureau of the China
Railway Construction Co.





\end{document}